 \documentclass[final,5p,times,twocolumn,authoryear]{elsarticle}

\usepackage{amssymb,amsthm,amsmath}
\usepackage{lipsum}
\usepackage{mathrsfs}

\begin{document}




\title{Compact Objects in close orbits as Gravitational Wave Sources: Formation Scenarios and Properties}



\author[0000-0002-1421-4427]{Zhenwei Li}
\affiliation{Yunnan Observatories, Chinese Academy of Sciences, Kunming, 650216, People's Republic of China}
\affiliation{Key Laboratory for the Structure and Evolution of Celestial Objects, Chinese Academy of Science, Kunming, 650216, People's Republic of China}
\affiliation{International Centre of Supernovae, Yunnan Key Laboratory, Kunming, 650216, People's Republic of China}

\author{Xuefei Chen}
\affiliation{Yunnan Observatories, Chinese Academy of Sciences, Kunming, 650216, People's Republic of China}
\affiliation{Key Laboratory for the Structure and Evolution of Celestial Objects, Chinese Academy of Science, Kunming, 650216, People's Republic of China}
\affiliation{International Centre of Supernovae, Yunnan Key Laboratory, Kunming, 650216, People's Republic of China}

\begin{abstract}
Gravitational Waves (GWs) provide a unique way to explore our Universe. The ongoing ground-based detectors, e.g., LIGO, Virgo, and KAGRA, and the upcoming next-generation detectors, e.g., Cosmic Explorer and Einstein Telescope, as well as the future space-borne GW antennas, e.g., LISA, TianQin, and TaiJi, cover a wide range of GW frequencies {from $\sim 10^{-4}\;\rm Hz$ to $\sim 10^3\;\rm Hz$} and almost all types of compact objects in close orbits serve as the potential target sources for these GW detectors. The synergistic multi-band GW and EM observations would allow us to study fundamental physics from stars to cosmology. {The formation of stellar GW sources has been extensively explored in recent years, and progress on physical processes in binary interaction has been made as well. Furthermore, some studies have shown that the progress in binary evolution may significantly affect the properties of the stellar GW sources.} 
In this article, we review the formation channels of compact objects in close orbits and discuss their implications for GW observations.
\end{abstract}



%
%
%





\section{Introduction}
\label{sec:1}
The first direct detection of gravitational wave (GW) merger event in 2015 opened a new window for us to explore the universe \citep{Abbott2016a}. A variety of stellar mergers have been reported by the LIGO-Virgo-KAGRA (LVK) collaboration, including binary black holes (BBHs), double neutron stars (DNSs), and black hole (BH)+neutron star (NS) binaries \citep{Abbott2016b,Abbott2017b,Abbott2019a,Abbott2019b,Abbott2021a,Abbott2023a,Abbott2023b,Abbott2023c}. The GW merger events have provided an excellent opportunity for studying the frontier science in modern astrophysics, such as stellar physics, binary interaction, and also shedding light on the cosmological evolution \citep{Abbott2018}. One of the most interesting GW events is the GW170817, the only source with the synergistic observations of GW and electromagnetic (EM) detectors \citep{Abbott2017b,Abbott2017c,Abbott2017d,Cowperthwaite2017,Savchenko2017,Troja2017,Smartt2017,Goldstein2017}, which opens the long-waited multi-messenger astronomy. The separate redshift measurement for this GW event makes it possible to use GW sources as {standard sirens} \citep{Abbott2017f}. Also, the accurately determined binary parameters of GW170817 allow us to probe the physics in extremely dense conditions \citep{Annala2018}. Although significant progress has been made since the successful detections of GW mergers, it also brings new challenges to theoretical studies in massive binary evolution. For example, we still do not know the individual formation channel for the GW events in LVK collaboration and the formation scenarios for several exotic GW mergers, e.g., BH in the pair-instability supernova (PISN) gap -- GW190521 \citep{Abbott2020a,Abbott2020b}, extreme mass ratio event -- GW190814 \citep{Abbott2020c}, etc. The upcoming next-generation GW detectors, e.g., Cosmic Explore \citep{Reitze2019} and Einstein Telescope \citep{Punturo2010}, possess the ability to find BBHs throughout the cosmic history. The GW merger samples are expected to be {enlarged by about $\sim 100000$} \citep{Hall2019}. The growing set of GW merger events unquestionably improves our understanding of the underlying populations and the basic physics. 




The ground-based GW detectors are designed to detect GW signals with frequencies larger than $\sim 10\;\rm Hz$. There are abundant GW sources that emit GW signals in lower frequencies, e.g., double white dwarf (DWD), AM Canum Venaticorum (AM CVn), ultra compact X-ray binary (UCXB), and also the inspiral BBH, DNS, and BH+NS. The space-borne GW antennas, such as DECIGO \citep{DECIGO2011}, LISA \citep{LISA2017,LISA2023}, TianQin \citep{Luoj2016}, and Taiji \citep{Ruan2020}, which cover the frequency band around $10^{-4}-1\;\rm Hz$, are proposed to detect such GW signals. For this reason, the space-borne GW detector is an indispensable element in the multi-messenger astronomy \citep{LISA2023}. At first, the multi-band GW observations with the combination of space-borne and ground-based GW detectors are expected to provide important implications for the formation scenarios of massive BBHs \citep{Sesana2016,Breivik2016}. Besides, the optical observations in combination with the LISA will find $\sim 100$ DWDs \citep{Korol2017,Lamberts2019,lizw2020}, the precisely determined binary parameters and distance are supposed to put a constraint on the Galaxy structure \citep{Korol2019}. Moreover, the GW detection of AM CVn and UCXBs can greatly improve our knowledge of the accretion physics with compact objects \citep{Breivik2018,Tauris2018}.

Forming compact binaries as GW sources via binary interaction is very complicated, which involves some essential but uncertain physical processes. For single stellar evolution, including the massive stellar winds, supernova (SN) explosion mechanism, natal kick of NS and BH, etc. For binary stellar evolution, including mass transfer stability, common envelope (CE) ejection, mass loss manners, etc. All of them remain controversial. In this article, we review the theoretical models in addressing the formation of stellar GW sources with an emphasis on the isolated binary evolution. 
In Section~\ref{sec:2}, we give a brief introduction to the single stellar evolution, and the main binary interaction processes are addressed in Section~\ref{sec:3}. The evolutionary routes to the detached binaries (e.g., DWD, NS+WD, DNS, BH+NS, BBH, and other potential GW sources) and accreting binaries (e.g., cataclysmic variable (CV), AM CVn, and UCXB) as GW sources are reviewed in Section~\ref{sec:4} and Section~\ref{sec:5}, respectively. Section~\ref{sec:6} discusses the significance of compact objects in close orbits for GW observations. A summary and outlook are given in Section~\ref{sec:7}.

\section{Single star evolution}
\label{sec:2}
The evolutionary fate of a single star strongly depends on two basic parameters: initial stellar mass and metallicity. The combination of the two parameters determines the critical stellar evolution processes, such as nuclear reaction, convection instability, stellar wind, etc. \citep{Eggleton2006,Kippenhahn2013}. In addition, rotation is another an important parameter that would alter the stellar structure. Remarkably, the inner elements can be effectively mixed for a massive star with rapid rotation, resulting in the so-called chemically homogenous evolution (CHE, \citealt{Maeder1997,Maeder1998,Maeder1999,Maeder2000,Meynet2000,Meynet2005}). This section will briefly introduce the main physical processes during stellar evolution.

\subsection{Single star evolutionary fate}
\label{sec:2.1}
According to the evolutionary products and the associating physical processes, we distinguish three mass regimes for stars at solar metallicity, i.e., $M \lesssim 8M_\odot$, $8M_\odot\lesssim M\lesssim 12M_\odot$, and $M\gtrsim 12M_\odot$. It should be noted that the boundaries for the mass regimes are far from being determined due to the poor understanding of some essential evolution stages, such as wind mass loss of massive stars, convective boundary mixing, and SN explosion mechanisms.  

Star with $M\lesssim 8M_\odot$ has a degenerate CO core at the AGB stage. Then most of the envelope material would be lost due to the thermal pulse instability caused by double-shell burning, leaving behind a CO WD finally \citep{Iben1983,Vassiliadis1993,Han1994,Herwig2005,Marigo2008,Hofner2018}. 

\begin{figure*}
	\centering 
	\includegraphics[width=\textwidth]{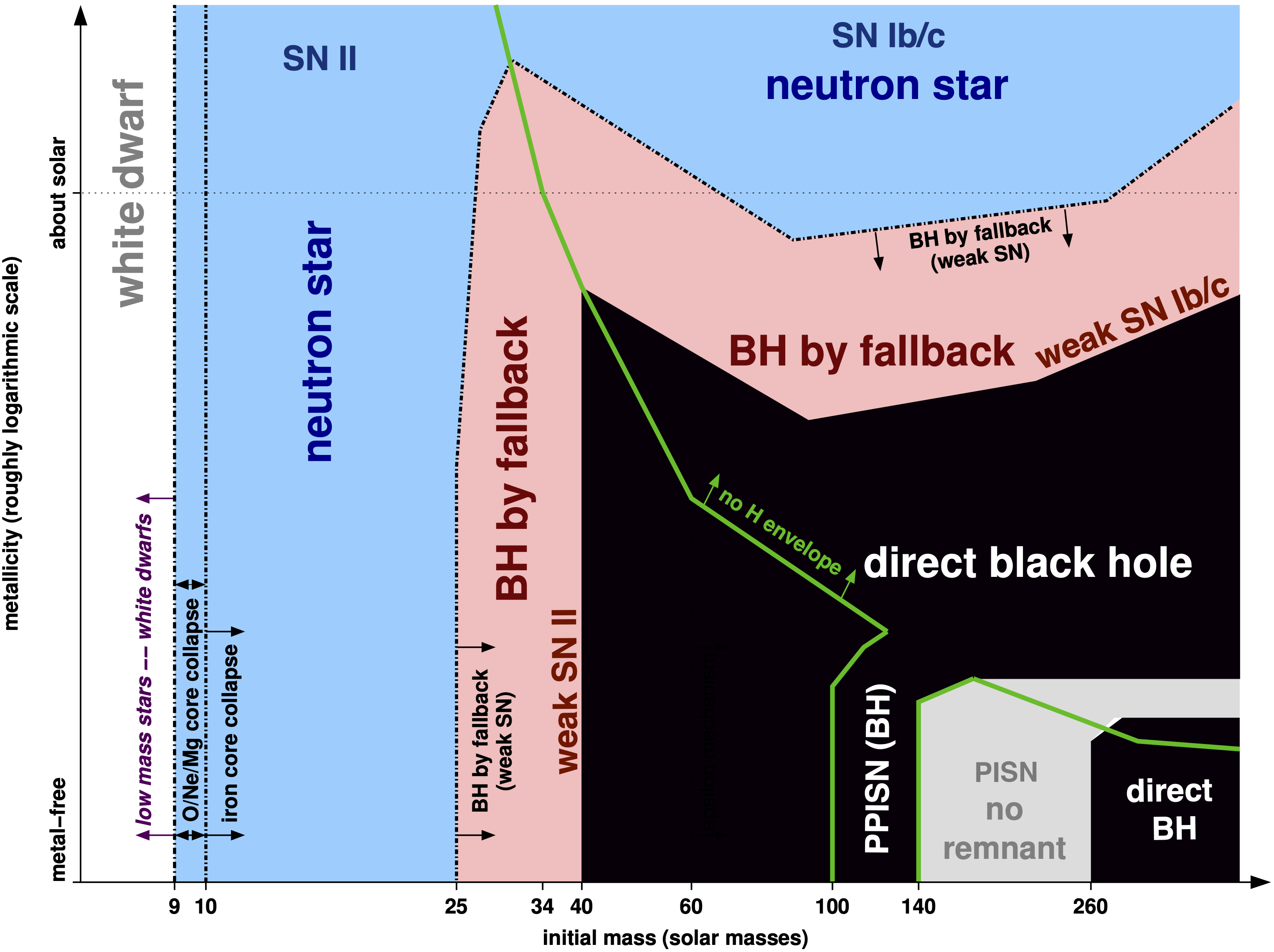}	
    \caption{Example of the standard expectations for the fates of massive stars as a function of initial mass and metallicity. The colored regions indicate the domain where various SN may occur. The white areas in the middle and lower right are for the direct BH formation. It should be noted that boundaries for those SN explosions are highly uncertain, and some values are different from the recent numerical simulations (e.g., \citealt{Sukhbold2016,Sukhbold2018,Ebinger2019}). Abbreviations: black hole, BH; supernova, SN. {Adapted from \citet{Heger2003}, reproduced by permission \textcopyright\ AAS.}} 
	\label{fig:1}%
\end{figure*}

For a star with $8M_\odot\lesssim M\lesssim 12M_\odot$, the CO core at AGB is not degenerate and can be ignited to form an ONe core. In the lower part of this mass regime ($\sim 8 - 10M_\odot$), the ONe core cannot be increased to the Chandrasekhar mass limit ($\sim 1.38M_\odot$) due to AGB wind and leaving an ONe WD finally \citep{Podsiadlowski2004,Woosley2015,Doherty2017}. In the mass range between $\sim 10-12M_\odot$, the ONe core would reach the Chandrasekhar mass limit; the high density of the core then leads to the electron captured by the magnesium, neon, and oxygen nuclei \citep{Nomoto1984a,Nomoto1987}. This process would release a significant part of electron degeneracy pressure, and the ONe core inevitably collapses until the NS is formed to defend the gravity. The so-called electron-capture SN (ECSN) is supposed to happen in this phase \citep{Nomoto1984a,Nomoto1987,Siess2007,Woosley2015,Jones2016,Zha2019}. NSs produced in electron-capture SNe have similar masses about $\sim 1.25M_\odot$ \citep{Wangb2020}. In a binary system, the ONe WD can grow its mass by accreting material from the companion star and produce an NS via the ECSN process (known as accretion-induced collapse, AIC; \citealt{Nomoto1984b,Nomoto1991,Woosley1992}). Although ECSN has not been confirmed in the observations and also there are some debates in theory, it is deemed to be an important formation channel of NSs (e.g., \citealt{Yoon2004,Dessart2006,Tauris2013,Leung2019,Wangb2020,Zha2022}).

For a more massive star with $M\gtrsim 12M_\odot$, the nuclei fuel continues to burn until the iron core is formed. The collapsed iron core finally leaves an NS or BH accompanied by the core collapse SN (CCSN; \citealt{Woosley2002,Heger2003,Smartt2009a,Janka2012,Fryer2012,Eldridge2004}). Stars with $12M_\odot\lesssim M\lesssim 20-25M_\odot$ generally produce NSs. The NSs produced from CCSNe show a large mass dispersion in the range of $\sim 1.1-2.2M_\odot$ since the iron core collapse process is not directly related to the Chandrasekhar mass limit but the physical details of the collapsing iron core (e.g. \citealt{Janka2012,Janka2016,Sukhbold2016,Sukhbold2018}). Therefore, the lower and upper limits, i.e., minimum and maximum NS mass limits, depend on the NS equation of state and are still in debate \citep{Lattimer2012,Ozel2016,Oertel2017}. For a star more massive than $\sim 20-25M_\odot$, the CCSN is followed by the fallback of material, resulting in the birth of BH. 

The above discussions mainly focus on the stars with solar metallicity. For stars in low-metallicity environments, the evolutionary fates would be significantly changed. Above $40M_\odot$, the low-metalicity stars may form black holes directly so that the extra mass loss due to the SN can be avoided. If stars have sufficiently low metallicity and the produced He cores have masses approximately in the range of $\sim 40-65M_\odot$, the pulsational PISN (PPISN) may happen, where violent pulsational instability is not energetic enough to disrupt the entire star, and the collapse continues until the formation of a BH \citep{Woosley2017,Marchant2019}. For very massive stars with $M\gtrsim 110M_\odot$ (He cores in the range of $\sim 64-130 M_\odot$), the pair-instability could trigger an explosive ignition of oxygen and lead to the total disruption of the star, i.e., PISN, finally leaving no remnant \citep{Barkat1967,Bond1984}. Stars with He cores more massive than $\sim 130M_\odot$ perform a direct collapse. Therefore, the theoretical models predict a gap in the BH mass spectrum, which ranges from $\sim 60M_\odot$ to $\sim 130M_\odot$ \citep{Heger2003,Belczynski2016,Spera2017,Stevenson2019}.

\begin{figure}
	\centering 
	\includegraphics[width=\columnwidth]{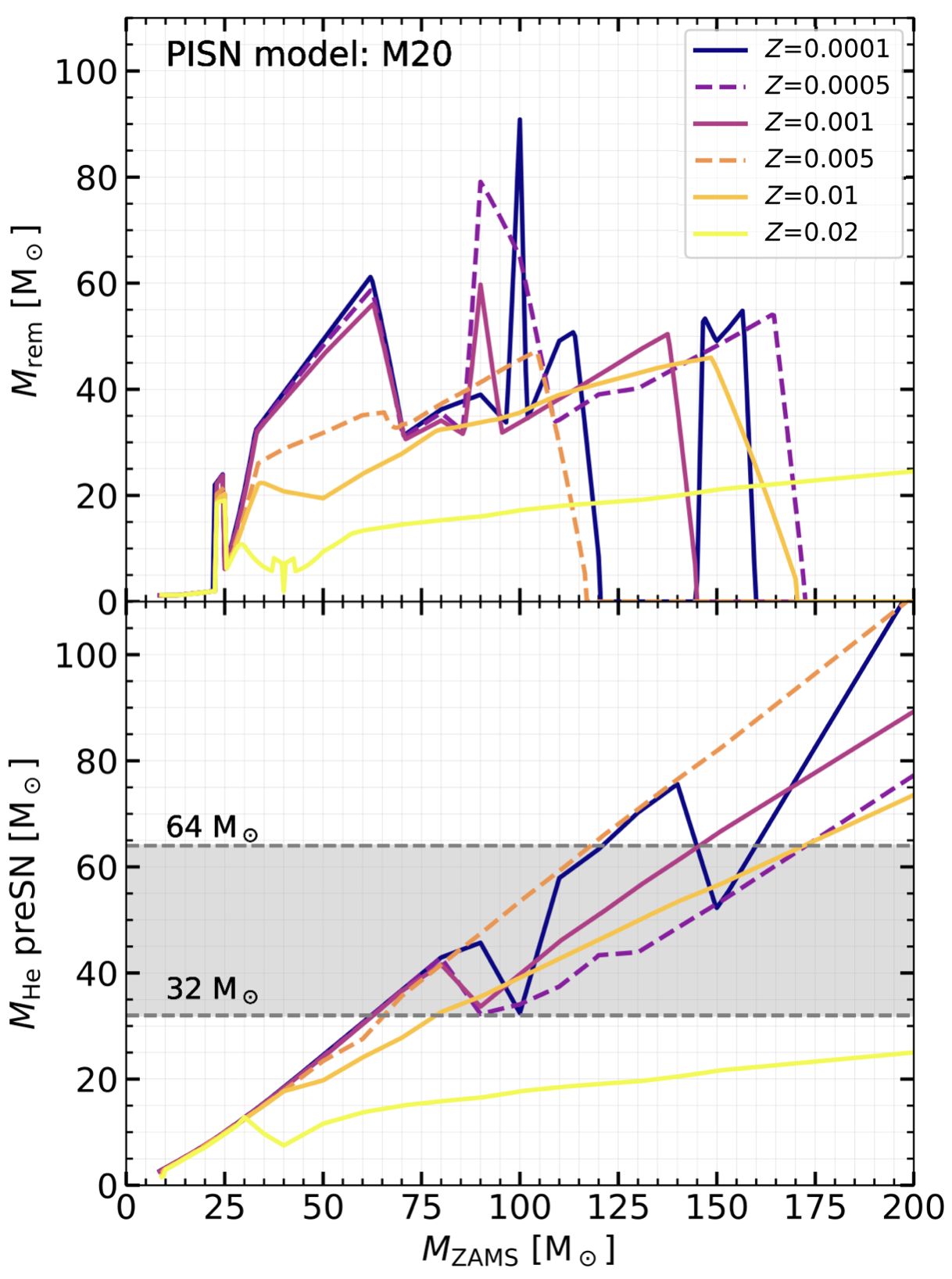}	
    \caption{Compact remnant mass and pre-SN He core masses from single stellar evolution as a function of the initial mass $M_{\rm ZAMS}$ based on the \texttt{PARSEC} stellar tracks with overshooting parameter of $\lambda_{\rm ov}$ = 0.5 \citep{Bressan2012}. The upper panel show the mass of the compact remnant with pair-instability \citet{Mapelli2020}. The lower panels show the pre-SN mass of the He core. The dashed horizontal lines mark the fundamental mass thresholds for the PISN models. The star undergos PPISN between $32\leq M_{\rm He}\leq 64M_\odot$ and explodes as a PISN for $M_{\rm He}>64M_\odot$. Adapted from \citet{Iorio2023}, {reproduced by permission \textcopyright\ RAS}.} 
	\label{fig:2}%
\end{figure}

The evolutionary fates for massive stars as a function of initial mass and metallicity are summarised in Figure~\ref{fig:1}. We stress that the boundaries of the initial masses for those SN explosions are highly uncertain, which strongly depends on the specific assumptions (e.g., \citealt{Sukhbold2016,Sukhbold2018,Ebinger2019}). Nevertheless, we could take a general picture of the final remnants for a given star with known mass and metallicity. It is important to note that in the low-metallicity environments, stars with initial mass $\gtrsim 40M_\odot$ form BHs directly. The directly collapsed process avoids mass loss due to the SN explosion, so we could expect massive BHs to form finally. {In Figure \ref{fig:2}, we present the compact remnant mass and pre-SN He core masses from single stellar evolution as a function of the initial mass, $M_{\rm ZAMS}$. Different implements, such as stellar wind, overshoot parameters, and PISN models, would significantly affect the results \citep{Iorio2023}. It should be noted that the final fates of massive stars, forming a NS after SN explosion or forming a BH directly, do not depend on the initial stellar mass monotonically, even for the stars with the same initial metallicity. Their final fates are decided by the CO core mass and the central C mass fraction at the end of core-He burning \citep{Chieffi2020,Patton2020,Schneider2021,Schneider2023}.} More detailed introductions about massive stellar evolution can be referred to in the recent reviews \citep{Vink2022,Costa2023,Tauris2023}.

\subsection{Stellar wind}
\label{sec:2.2}
Stellar wind is a critical factor for understanding the stellar evolution products, which determines the final He core mass and the pre-SN He star mass. There are several types of winds with different driven mechanisms, such as Alfven waves-driven and dust-driven for cool stars (e.g. RGB and AGB stars; \citealt{deJager1988,Vassiliadis1993,Vassiliadis1994,Hofner2007,Hofner2018}) and line-driven for Massive hot stars (e.g. OB stars and WR stars; \citealt{Nugis2000,Smith2014,Smartt2009b,Langer2012}). Despite significant theoretical and observational progress, wind mass loss remains an important open question in stellar physics (e.g., \citealt{Decin2021,Vink2022}). It is commonly accepted that stellar wind has a strong correlation with the metallicity \citep{Maeder1992,Vink2001,Muijres2011,Muijres2012}: $\dot{M}\propto Z^{0.85}$ (but see \citealt{Vink2021}, who found the $Z$-dependence of the mass-loss rate can be as shallow as $\dot{M}\propto Z^{0.42}$). We take solar metallicity, for example, to illustrate the main wind mass-loss processes for stars in several typical mass ranges. 

For a low- and intermediate-mass star ($\lesssim 8M_\odot$), there is considerable mass loss as the star ascends on the giant stage (RGB and AGB). The typical mass loss rates are about $10^{-10}-10^{-7}M_\odot\;\rm yr^{-1}$ for stars at RGB and $10^{-7}-10^{-4}M_\odot\;\rm yr^{-1}$ for stars at AGB \citep{Suzuki2007,Hofner2018,Yasuda2019}. At the thermal pulse AGB, most of the envelope is lost and leaving CO or ONe WD finally \citep{Schoenberner1983,Kippenhahn2013}. Unlike low- and intermediate-mass stars, stars massive than $8M_\odot$ experience strong wind mass-loss at earlier stages. The radiatively driven wind can efficiently remove the hydrogen-rich envelope and alter the remnant helium (He) core mass \citep{Sukhbold2016,Sukhbold2018}. The high luminosity makes the winds stronger with more massive stars (See \citealt{Vink2022} for the recent review). 

\begin{figure}
	\centering 
	\includegraphics[width=\columnwidth]{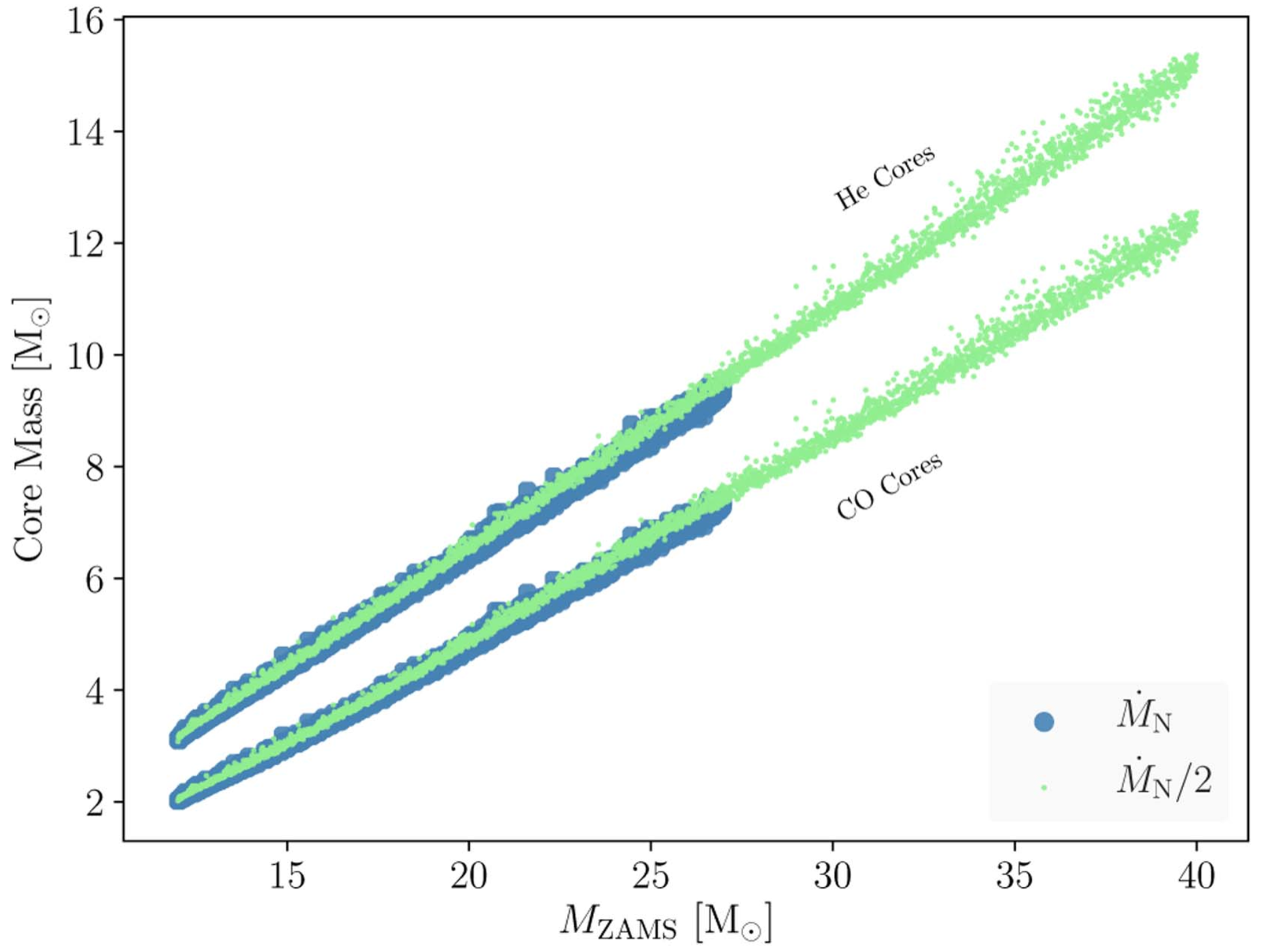}	
    \caption{He and CO core masses for pre-SN stars as a function of initial stellar mass. $\dot{M}_{\rm N}$ and $\dot{M}_{\rm N}/2$ mark the standard mass loss rate and half of $\dot{M}_{\rm N}$ adopted in \citet{Sukhbold2018}. It is clear that the final helium and CO cores are well determined by the star’s initial mass without respect to the wind mass loss rate. Adated from \citet{Sukhbold2018}, {reproduced by permission \textcopyright\ AAS.}} 
	\label{fig:3}%
\end{figure}

Stars massive than $\sim 25M_\odot$ will not become red supergiants (RSGs; the lack of luminous RSGs in the observations is known as Humphreys-Davidson limit; \citealt{Humphreys1979}), the Wolf-Rayet stars may be formed when most of the hydrogen-rich envelope is stripped by the stellar wind (the surface hydrogen mass fraction $X_{\rm s}\lesssim 0.4$; artificially defined based on the observations; \citealt{Crowther2007,Smith2014}). If the star mass is not more massive than $\sim 40M_\odot$, the wind mass is not strong enough to directly dig into the He core. Therefore, there is an approximately linear correlation between the He core mass and the ZAMS mass \citep{Sukhbold2018,Woosley2019}, as shown in Figure~\ref{fig:3}. WR stars show robust and broad emission lines in the spectrum and generally have stronger wind mass-loss rates than the same luminosity O-star winds \citep{Vink2002,Crowther2007,Woosley2019}. 

More massive stars can be driven to the luminous blue variables (LBVs) phase where almost all hydrogen in the envelope has been depleted during the core hydrogen burning phase \citep{Heger2003,Smith2014}. The eruptive mass-loss of LBVs can dramatically change the star structure in a short time \citep{Smith2006}. However, due to the infrequent eruption events, the physical mechanism during the LBV phase still needs to be better understood \citep{Massey2007,Langer2012,Kalari2018}. If stars with masses of $80-100M_\odot$ or more massive, the luminosity may exceed the Eddington limit \citep{Grafener2011,Vink2011}. The stars may avoid the LBV phase and directly evolve to the luminous WNH stars, i.e., H-rich WR stars \citep{Kudritzki2000,Crowther2007,Puls2008,Langer2012,Sen2023}. 

\begin{figure}
	\centering 
	\includegraphics[width=\columnwidth]{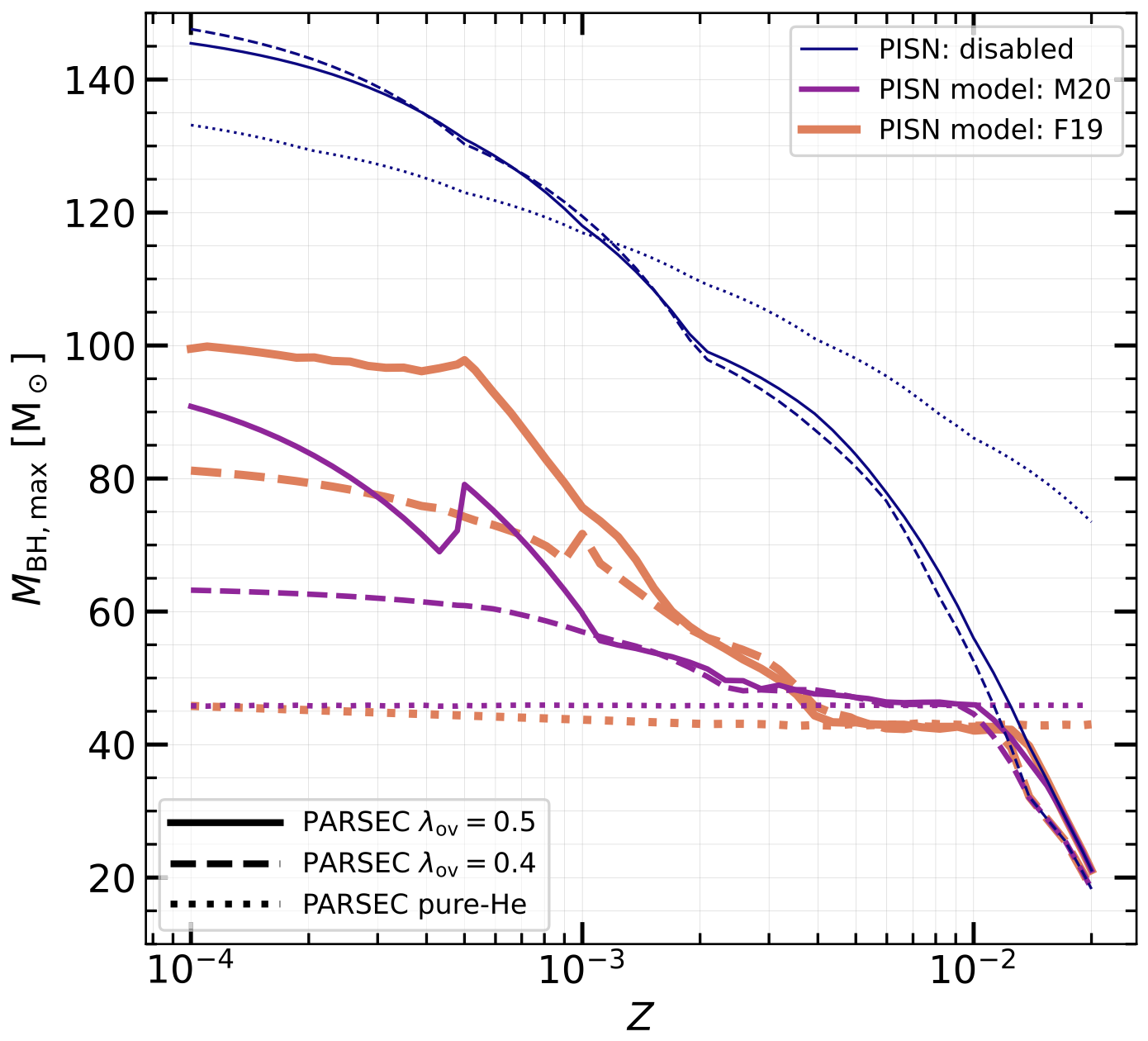}	
    \caption{Maximum BH mass from single stellar evolution as function of metallicity. The initial ZAMS stars are calculated up to $200M_\odot$.  
    The rapid explosion model of \citealt{Fryer2012} is adopted in calculating CCSN. For pair instability models, no pair-instability correction: very thin blue lines; M20 (based on \citealt{Mapelli2020}): thin violet lines; F19 (based on \citealt{Farmer2019}): thick orange lines. Solid and dashed lines are for models with different overshoot parameters, and dotted lines are for the pure-He stars. Adated from \citet{Iorio2023}, {reproduced by permission \textcopyright\ RAS}.} 
	\label{fig:4}%
\end{figure}

The stellar wind mass loss will be largely depressed at low-metallicity environment, as introduced above. Then the massive remnants can be expected. Figure~\ref{fig:4} presents the maximum BH mass from single stellar evolution as function of metallicity. The results are taken for \texttt{SEVN} code \citep{Iorio2023}. We see that the maximum BH mass for $Z=0.02$ is about $25M_\odot$, while it becomes $\sim 60-100M_\odot$ for $Z\lesssim 0.0005$. It is remarkable that \citealt{Bavera2023} adopting self-consistent stellar wind description found that star at solar metallicity can produce BHs with masses beyond $30M_\odot$. Nevertheless, low metallicity is indispensable to explain the observed BBH populations in LVK collaboration (see Section~\ref{sec:4.4} and \ref{sec:6.1} for more details).

\subsection{Natal kick}
\label{sec:2.3}
The natal kicks of NS and BH are crucial in massive binary evolution. The evidence of SN kick arise from the large dispersion of pulsars' space distributions around the Galaxy plane, while the progenitor stars are concentrated in the thin layer \citep{Gunn1970}. The large dispersions of pulsars in the space can only be explained by assuming that NSs received a large kick velocity as they were born (e.g., \citealt{Shklovskii1970,Taylor1977,Lyne1994}). \citet{Hobbs2005} fitted the proper motions of $73$ young pulsars in the Galaxy and found that the kick velocities of these pulsars follow a Maxwellian distribution with the mean value of $400\pm 40 \;\rm km\;\rm s^{-1}$ and one-dimensional root mean square $\sigma_{\rm kick}=265\;\rm km\;s^{-1}$. {Several pulsars are even found with kick velocity larger than $1000\;\rm km\;s^{-1}$, e.g. IGR J11014-6103, PSR J0002+6216, and B1508+55  \citep{Cordes1993,Hobbs2005,Chatterjee2005,Tomsick2012,Pavan2014,Schinzel2019}.}

In a theoretical view, the natal kick arises from an asymmetric SN explosion when the central compact object (NS or BH) is born. Several mechanisms may drive the kicks, such as hydrodynamical driven and neutrino driven \citep{Lai2000,Lai2001,Wangc2006,Fryer2006,Ott2006,Kotake2012,Janka2012,Janka2016,Wongwathanarat2013,Janka2017,Muller2019,Burrows2019}. The kick strongly depends on the SN explosion mechanisms. In general, higher natal kicks can be expected for explosions of more massive SN progenitors \citep{Arnett2011,Wongwathanarat2013,Gessner2018}. \citet{Janka2017} adopted the gravitational tug-boat mechanism in asymmetric mass ejection of neutrino-driven explosions and found a simple analytic expression for the kick velocity regarding explosion energy, ejecta mass, and asymmetry of SN explosion. For the ECSN, the electron-capture process is supposed to be rapid, and also due to the relatively lower explosion energy ($\lesssim 10^{50}\;\rm erg$), the kicks are expected to be small with typical kick velocity of $<50\;\rm km\;s^{-1}$ \citep{Podsiadlowski2004,Kitaura2006,Dessart2006,Janka2012,Wongwathanarat2013,Gessner2018,Stockinger2020}. Stars with masses range from $\sim 12M_\odot$ to $\sim 25M_\odot$ may produce NSs via CCSN \citep{Sukhbold2016,Muller2018,Ebinger2019}. Comparing with ECSN, the typical explosion energy of CCSN is above $10^{51}\;\rm erg$, therefore we could expect a large natal kick for NS from CCSN (e.g. \citealt{Janka2017,Ebinger2019,Ebinger2020}).

For the natal kicks of newly formed BHs, there are poor constraints due to the rare detections of BHs in the EM observations (e.g., \citealt{Brandt1995,Bailyn1995,Nelemans1999,Remillard2000,Mirabel2001,Willems2005,Gualandris2005,McClintock2006,Corral2016,Remillard2006,Dhawan2007,Fragos2009,Mandel2016,Repetto2017,Chauhan2019,Atri2019,Lam2022,Mroz2022,Sahu2022,Elbadry2023a,Elbadry2023b,Sukanya2023}). 
{it is not obvious if the natal kicks of BHs follow the same distribution as that of NSs. 
If the natal kicks of BHs are also driven by the netrino-based mechanism, similar to the NSs, then it predicts that the natal kicks will be inversely proportional to the mass of a BH \citep{Janka2013,Rodriguez2016b,Wiktorowicz2019}. However, such a correlation between the natal kicks and BH masses are not found in the observations \citep{Repetto2012}. For example, GS 2023+338 \citep{Miller2009} with BH mass of $9.0 \pm 0.2$ has a small kick velocity ($< 45 \;\rm km\; s^{-1}$), while a more massive BH ($M_{\rm BH}=10.2\pm 1.5M_\odot$) in XTE J1819-254 \citep{Repetto2012,Belczynski2016b} has significant natal kick ($>100\;\rm km\; s^{-1}$). An alternative model proposed for the natal kicks of BHs scales the BH kick magnitude via the fallback material \citep{Fryer2012}, i.e. 
\begin{eqnarray}
  w_{\rm BH}/w_{\rm NS} = 1-f_{\rm fb},
\end{eqnarray}
where $w_{\rm BH}$ and $w_{\rm NS}$ are the kick velocities of BH and NS, $f_{\rm fb}$ is the fraction of ejected mass falling back onto a compact object. This model produces large natal kick of a BH comparing to the NS with a less fallback, and also predicts low natal kicks for BH formed with noticeable fallback. The result is boardly in agreement with observational estimates for BH X-ray binaries (e.g., \citealt{Belczynski2016b,Repetto2017}).}

As an essential input parameter in the binary population synthesis, the kick velocity strongly affects the merger rates of double compacts (e.g., \citealt{Belczynski2016,Kruckow2018,Breivik2020a,Spera2019}). However, we must admit that the nature of the natal kicks for SN remnants is still an open question. A deep understanding of the SN explosion mechanisms may illuminate this issue (e.g., \citealt{Janka2012,Janka2013,Janka2016,Sukhbold2016,Sukhbold2018}).

\section{Binary star evolution}
\label{sec:3}
In a binary system, if the binary separation is wide enough, the stars will not be affected by each other, so one can describe the evolution of these two stars via single stellar evolution theory. If the binary separation is close, the evolution processes for a binary show significant differences. Binary interactions such as tides, mass transfer, and wind accretion would lead to some unique products that cannot be formed from single evolution e.g., SNe Ia, extremely low-mass WDs, and hot subdwarfs, etc. (\citealt{Iben1984,Han2002,Han2003,Han2004,Heber2009,Wangb2012,Heber2016,Han2020,Liuz2023} and references therein). In this section, we briefly introduce the commonly adopted binary evolution model and the key physical processes during binary evolution.

\subsection{Roche lobe model}
\label{sec:3.1}
The Roche lobe model is a widely accepted model to describe the equipotential surfaces of two stars in a co-rotation and circular orbit. The equipotential surfaces with inner Lagrangian point $L_1$ are the Roche lobes of the two stars. If a star evolves to fill its Roche lobe, the unbound material is transferred to the companion star via the $L_1$ point. This process is referred to as Roche lobe overflow (RLOF), the most important mode for the mass transfer between two binary components. It is convenient to describe the Roche lobe in a spherical-equivalent approximation, i.e., \citep{Eggleton1983}
\begin{eqnarray}
  R_{{\rm RL},j} = \frac{0.49q_{j}^{2/3}a}{0.6q_{j}^{2/3}+\ln \left(1+q_{j}^{1/3}\right)},
\end{eqnarray}
where $j$ is the index identifying each star, $q_{\rm 1} = M_{1}/M_2$ and $q_2 = M_2/M_1$, the accuracy is within $1\%$ for all values of $q$. Depending on the binary separation and the companion radius, the transferred material may directly be accreted onto the companion surface or form an accretion disk \citep{Lubow1975,Ulrich1976,Thomas1977,deMink2013,Marchant2016}. Although binary evolution plays a key role in modern astrophysics, there are still many uncertainties in theory, e.g., the criteria of dynamically unstable mass transfer, CE ejection, and the non-conservative mass transfer \citep{Chenx2024}.

\subsection{Mass transfer stability}
\label{sec:3.2}
The mass transfer stability is a longstanding problem crucial for determining a binary system’s evolution product. Whether or not the mass transfer is dynamically stable is often understood in terms of the response of donor star radius and Roche lobe radius to the mass loss (e.g., \citealt{Webbink1985,Hjellming1987,Soberman1997,Tout1997,Chenx2003,Chenx2008,Pavlovskii2015}). If the mass loss is slow enough that the star remains in thermal equilibrium, the radial response of the donor to mass loss is defined as
\begin{eqnarray}
  \zeta_{\rm eq} = \left(\frac{d\;\ln R_1}{d\;\ln M_1}\right)_{\rm eq},
\end{eqnarray}
where $R_1$ and $M_1$ are the donor radius and mass, respectively. If the mass loss is rapid enough that the donor is out of thermal equilibrium but still retains hydrostatic equilibrium, then the radial response is given by
\begin{eqnarray}
  \zeta_{\rm ad} = \left(\frac{d\;\ln R_1}{d\;\ln M_1}\right)_{\rm ad}.
\end{eqnarray}
The Roche lobe response to the mass loss is defined as
\begin{eqnarray}
  \zeta_{\rm RL} = \frac{d\;\ln R_{\rm RL,1}}{d\;\ln M_1}.
\end{eqnarray}
If a Roche lobe filling star satisfies $\zeta_{\rm eq}>\zeta_{\rm RL}$, the donor remains inside its Roche lobe with retaining thermal equilibrium. Then, the mass transfer phase is driven by the radial expansion due to the nuclear burning. Binary mass transfer during this phase is known as nuclear timescale mass transfer. In the case of $\zeta_{\rm ad}>\zeta_{\rm RL}>\zeta_{\rm eq}$, the star is out of thermal equilibrium but still retaining hydrostatic equilibrium. The thermal expansion of the donor drives the mass transfer. Such a phase is known as thermal timescale mass transfer. In the extreme case of $\zeta_{\rm RL}>\zeta_{\rm ad}$, the star will depart from hydrostatic equilibrium, and the mass transfer will proceed on a dynamical timescale. The donor star’s rapid expansion will soon engulf the companion star, and the binary enters into the CE phase \citep{Paczynski1976}. The tangency condition of $\zeta_{\rm ad}=\zeta_{\rm RL}$ defines a critical mass ratio, $q_{\rm c}$, above which the mass transfer is unstable to dynamical timescale mass transfer \citep{Hjellming1987,Tout1997,Hurley2002}.

In the early studies, the critical mass ratio is obtained based on the polytropic stellar models with the power-law equation of state, i.e., $p\approx \rho^{(1+1/n)}$, where $p$ is the pressure, $\rho$ is density and $n$ is the polytropic index \citep{Hjellming1987,Webbink1988}. Three cases are often considered, i.e., complete polytrope of $n=3$ for MS stars, composite polytrope of $n=3$ cores with $n=3/2$ envelopes for giant stars, and condensed polytrope of $n=3/2$ for WDs. Assuming conservative mass transfer, it approximately gives $q_{\rm c} \sim 3$ for an MS star and $q_{\rm c}=2/3$ for a degenerate WD. The $q_{\rm c}$ for a giant star with degenerate core is given by \citep{Hjellming1987}
\begin{eqnarray}
  q_{\rm c} = 0.362+\frac{1}{3(1-M_{\rm c}/M_1)},
\end{eqnarray}
where $M_{\rm c}$ is the core mass of the giant. The results of $q_{\rm c}$ based on polytropic models introduced above are widely used in binary population synthesis (BPS) studies \citep{Han2020}. 

Subsequent works with detailed binary evolution calculations argue that the criterion based on polytropic models is not appropriate \citep{Tauris1999,Podsiadlowski2002}. On the one hand, the details of dealing with the mass loss due to the non-conservative mass transfer wind should have important effects on the values of $q_{\rm c}$ (e.g., \citealt{Chenx2008}). On the other hand, the polytropic model is unsuitable for the realistic stellar with the developed core. \citet{Geh2010} constructed the adiabatic mass-loss models with considering the realistic stellar model. In the series works, \citet{Geh2015,Geh2020a} investigated the adiabatic mass loss sequences of Population I stars with masses ranging from $0.1M_\odot$ to $100M_\odot$ by covering their evolutionary stages from the ZAMS to the tip of AGB. An extraordinary result in their works is that the mass transfer tends to be more stable than previously believed for RGB and AGB stars. 
 
The mass transfer instability is also investigated by comparing the donor radius and the outer Lagrangian point $L_2$ (e.g., \citealt{Pavlovskii2015,Pavlovskii2017,Geh2020b}). The dynamical mass transfer is unavoidable when the radius of the donor overfills its outer Lagrangian point since the lost material from $L_2$ would carry away more specific angular momentum (relative to the specific angular momentum of the accretor) and leads to the dramatical shrinkage of the orbit. {Based on this method, \citet{Pavlovskii2015} found that the critical initial mass ratio for the donor star with a well-developed outer convective envelope varies from 1.5 to 2.2, which is about twice as large as previously believed. By using the new criteria, \citet{Pavlovskii2017} found that the predicted rates of ultra-luminous X-ray sources powered by a stellar-mass BH are high enough to explain the number of observed bright ultraluminous X-ray sources.}

Though large developments have been made during the last decades, the mass transfer stability is far from being determined due to the complex of rapid mass loss processes \citep{Geh2020a,Geh2023a}. A more detailed discussion about this issue can be found in \citet{Chenx2024}.

\subsection{Stable Roche lobe overflow}
\label{sec:3.3}
The binary mass exchange proceeds on a long timescale via the RLOF if the mass transfer is dynamically stable. The binary’s further evolution depends on the donor star’s structure and the orbit angular momentum loss. It is convenient to differentiate three types of mass transfer phases, i.e., Case A for donor filling its Roche lobe during core-hydrogen burning, Case B for donor filling its Roche lobe during shell-hydrogen burning, and Case C for donor filling its Roche lobe during He shell burning \citep{Kippenhahn1967}. In a particular case, the naked He stars with mass $\lesssim 3.2M_\odot$ would expand to large radii during the shell-He burning phase and can also initiate the mass transfer to its companion; such a phase is often called a Case BB mass transfer \citep{Savonije1976,DeGreve1977,Delgado1981,Tauris2006}. The ultra-stripped He core has important implications for NS’s birth (e.g. \citealt{Tauris2012,Tauris2015,Tauris2017,Jiang2021,Sawada2022,Yans2023,Richardson2023}).

Several types of angular momentum loss mechanisms affect the orbital evolution, mainly including GW radiation, magnetic braking, mass loss, and tides (e.g., \citealt{Hurley2002}). The rate of orbital angular momentum loss carried by the GW radiation is written as \citep{landau1975}
\begin{eqnarray}
  \frac{{\rm d}J_{\rm orb}}{{\rm d}t} = -\frac{32}{5}\frac{G^{7/2}M_1^2M_2^2\sqrt{M_1+M_2}}{c^5a^{7/2}},
\end{eqnarray}
where $c$ is the light speed in a vacuum, $G$ is the gravitational constant and $a$ is the binary separation. The GW radiation always leads to orbital shrinkage and dominates for binary with a very close orbit, such as AM CVn and UCXBs \citep{Solheim2010,Postnov2014,Chenh2021}.

The magnetic braking is caused by the loss of magnetic stellar winds for low-mass stars \citep{Huang1966,Mestel1968,Skumanich1972}. Several magnetic braking prescriptions are proposed based on the magnetic field geometry and winds (e.g., \citealt{Verbunt1981,Rappaport1983,Kawaler1988,Dekool1992,Sills2000}). One of the most commonly adopted MB prescriptions is the so-called Skurmanich law \citep{Skumanich1972,Rappaport1983}:
\begin{eqnarray}
  \dot{J}_{\rm MB} &=& -5.83\times 10^{-16}\frac{M_{\rm env}}{M_{\rm 1}} \\\nonumber
  & & \cdot\left(\frac{R_{\rm 1}\omega_{\rm spin}}{R_\odot\;\rm yr^{-1}}\right)^{\gamma_{\rm MB}}M_\odot R_\odot^2\rm yr^{-2},
\end{eqnarray}
where $\gamma_{\rm MB}$ is a free parameter with typical value in the range of $0-4$ \citep{Knigge2011}, $M_{\rm env}$ is the donor's envelope mass, and $\omega_{\rm spin}$ is the spin angular velocity. Although the Skurmanich law is widely applied in low-mass binary evolutions, there are difficulties in explaining some specific binary systems, such as the low-mass X-ray binary (LMXB) populations and millisecond pulsars in tight orbits (e.g., \citealt{Podsiadlowski2002,Istrate2014a}). Recently, \citet{Van2019b} proposed a new MB prescription, namely convection and rotation boosted (CARB) prescription, and the angular momentum loss due to MB is largely enhanced in comparison with the classical Skurmanich prescription (see also \citealt{Van2019a,Van2021}). The CARB prescription has been successfully applied in the populations of LMXB, AM CVn, and millisecond pulsar (MSP) binaries (e.g., \citealt{Van2021,Chenh2021,Dengz2021,Belloni2023a}).

The lost mass escaping from the binary system can also carry away the orbital angular momentum, leading to the shrinkage or widening of the orbit \citep{Tauris2006,Postnov2014}. According to the mass loss manners during the stable RLOF, the orbital angular momentum loss is given by \citep{Heuvel1994,Soberman1997,Tauris2006}
\begin{eqnarray}
  \frac{\dot{J}_{\rm ML}}{J_{\rm orb}} = \frac{\alpha+\beta q^2+\delta\gamma(1+q)^2}{1+q} \frac{\dot{M}_1}{M_1},
\end{eqnarray}
where $q =M_1/M_2$ is the mass ratio, $\alpha$ is the fraction of mass lost from the donor in the form of a direct fast wind, $\beta$ is the fraction of mass lost from the vicinity of the accretor, and $\delta$ is the fraction of mass lost from a circumbinary disk with a radius of $\gamma^2 a$. {An extra but recently found important angular momentum loss mechanism is that masses lose through the outer Lagrangian point. In this case, the lost material would carry away a significant part of angular momentum and lead to dramatically orbital shrinkage, which positively affects the formation of compact binaries in close orbits \citep{Marchant2021,Picco2023}. An essential issue is how many masses are lost through the outer Lagrangian point. In the recent simulations, \citet{Luw2023} proposed that, at a sufficiently high mass transfer rate (typically a few $10^{-4}M_\odot\;\rm yr^{-1}$), the accretion disc around the companion becomes geometrically thick (or advection-dominated) near the disc outer radius, which results in a large fraction ($\gtrsim 50\%$) of transferred mass lost through outer Lagrangian point. The effects of this result on binary evolution deserve to further investigate.}

The tidal friction will circularize the orbit and bring the stars into synchronized co-rotation. Tides strongly depend on the stellar structure  (convective or radiative envelope) and the orbital separation \citep{Zahn1975,Zahn1977,Hut1981,Hurley2002}. The tidal dissipation is more efficient for stars with convective envelopes (but see \citealt{Nie2017} who find the convective damping may be less effective than previously believed). For two degenerate stars, the tides only become important when binary separation is very small (see \citealt{Hurley2002} for more details). 

{Typically, if donor stars undergo stable RLOF with degenerate core, the remnants, such as hot subdwarf and WD, have a strong correlation to the orbital period, which is known as mass-orbital period relation \citep{Rappaport1995}. This relation arises from the core mass-radius relation for a giant, and then is independent of the angular momentum loss mechanism. The mass-orbital period relation has been confirmed with the discoveries of numerous long-period hot subdwarf binaries, blue stragglers, NS and He WD binaries, and DWDs \citep{Chenx2009,Linj2011,Chenx2013,Chenx2017,lizw2019,Gaos2023}. }

\subsection{Common envelope evolution}
\label{sec:3.4}
For the case of $\zeta_{\rm ad}<\zeta_{\rm RL}$, the binary mass transfer would be dynamically unstable, and the short timescale of the mass loss then leads to the emergence of CE. CE phase is deemed to be an indispensable process to produce close-orbit compact binaries, such as CVs, GW sources, SN Ia, where significant orbital shrinkage is required \citep{Ostriker1973,Webbink1975,Paczynski1976,Heuvel1976,Webbink1984}. The CE ejection process may involve many complicated physical processes, such as tidal drag force, element recombination, magnetic field, etc., making it one of the most important open questions in astrophysics. The commonly adopted phenomenological description based on energy conservation (known as $\alpha-$mechanism) is proposed to simplify the CE ejection process \citep{Heuvel1976,Webbink1984}. In this scenario, the release of orbital energy is used to eject the CE, and the CE can be ejected successfully when
\begin{eqnarray}
  \alpha_{\rm CE}|\Delta E_{\rm orb}|>|E_{\rm gr}+\alpha_{\rm th}E_{\rm th}|,
\end{eqnarray}
here $\alpha_{\rm CE}$ is the CE efficiency and is defined as the fraction of reduced orbital energy used in ejecting the CE. This parameter can be constraint by post-CE binaries in the observations (e.g., \citealt{Zorotovic2010,Demarco2011,Geh2022,Scherbak2023,Geh2023b}), but it is still unclear whether this is a common value available for all types of binaries \citep{Ivanova2013,Ivanova2020}. The values of CE efficiency is an essential parameter in the prediction of merger rates of compact binaries, and the results may vary by one or two orders of magnitude with different assumptions of $\alpha_{\rm CE}$ (see Section~\ref{sec:4} and \ref{sec:6} for more details). $E_{\rm gr}$ is the gravitational binding energy and $E_{\rm th}$ is the thermal energy of the envelope. $\Delta E_{\rm orb}$ is the released energy with orbital shrinkage and is written as
\begin{eqnarray}
\Delta E_{\rm orb} =  \frac{GM_2M_{\rm c}}{2a_{\rm f}}-\frac{GM_1M_2}{2a_{\rm i}},
\end{eqnarray}
where $M_{\rm c}$ is the core mass of the donor, $a_{\rm i}$ and $a_{\rm f}$ is the binary separation before and after the CE ejection, respectively. The binary energy $E_{\rm gr}$ and the thermal energy $E_{\rm th}$ are obtained based on the stellar structure, i.e.
\begin{eqnarray}
  E_{\rm gr} = \int_{M_{\rm c}}^{M_{\rm s}} -\frac{Gm}{r}{\rm d}m,
\end{eqnarray}
and
\begin{eqnarray}
  E_{\rm th} = \int_{M_{\rm c}}^{M_{\rm s}}U{\rm d}m,
\end{eqnarray}
where $M_{\rm s}$ is the stellar surface mass, $M_{\rm c}$ is the core mass, $U$ is the internal energy of the envelope \citep{Han1994}. One important point in calculating the envelope’s binding energy and thermal energy is the definition of the core boundary. This question is not trivial and would significantly affect the final products (e.g., \citealt{Ivanova2011a,Ivanova2011b}). Besides, the binding energy is changed during the spiral-in process. Therefore, problems may arise from only considering the envelope binding energy at the onset of CE phase. {The $\beta-$mechanism with taking into account the change of binding energy has been proposed by \citet{Geh2022,Geh2023b}, which seems to be more realistic in dealing with the CE ejection processes. }

An alternative prescription for the CE ejection process is $\gamma-$mechanism, based on the angular momentum balance \citep{Nelemans2000,Nelemans2005}. \citet{Nelemans2000} adopted $\gamma-$mechanism successfully reproducing some DWDs which cannot be explained by $\alpha-$mechanism (see also \citealt{Toonen2012,Toonen2017}). Nevertheless, there is no clear physical explanation for the $\gamma-$mechanism. In recent work, \citet{lizw2023} carried out BPS simulations on DWD populations and found that the stable non-conservative mass transfer may have an effect akin to the $\gamma-$description. Therefore, the non-conservative mass transfer with angular momentum loss may be the process underlying the $\gamma-$mechanism (See also \citealt{Woods2012,Ivanova2013,Postnov2014}).

\section{Detached Binary}
\label{sec:4}
Binary interaction could produce a variety of compact binaries. 
In this section, we focus on the formation scenarios of detached binaries as GW sources where both components are within the Roche lobes.

\subsection{DWD}
\label{sec:4.1}
DWDs are of great interest since they are the most common double compact objects in the Galaxy/Universe \citep{Tutukov1981,Iben1984,Iben1997,Han1998,Nelemans2001a,Toonen2017}. However, due to their intrinsic faintness, there are only $\sim 150$ DWDs are observed up to now (e.g., \citealt{Saffer1988,Marsh1995,Maxted1999,Badenes2009,Napiwotzki2003,Brown2010,Rebassa2019,Burdge2020a}). Most of DWDs are found including an extremely low-mass WD (ELM WD; He core WD with a typical mass of $\lesssim 0.25M_\odot$) companions (e.g., \citealt{Brown2010,Kilic2011,Brown2012,Kilic2012,Brown2013,Gianninas2015,Brown2016,Brown2020,Brown2022,Kosakowski2020,Burdge2020b,Wangk2022,Kosakowski2023a,Yuanh2023a,Yuanh2023b}), since ELM WDs can sustain high luminosity for a long timescale due to the massive H-rich envelope \citep{Althaus2013,Istrate2014b,Istrate2016,Chenx2017,lizw2019}.

Several channels may lead to the formation of DWDs, as investigated in \citet{Han1998}. In the most general scenario, the binary experiences two mass transfer phases to produce a DWD. The massive component evolves to fill its Roche lobe first for a primordial binary with two zero-age main sequence (MS) stars. The mass transfer may be stable or unstable according to the abovementioned mass transfer stability criterion. After the first (stable/unstable) mass transfer phase, the primary core is left and finally evolves to a WD. The unevolved/less-evolved companion subsequently fills its Roche lobe, leading to the occurrence of the second mass transfer phase. Similarly, the binary may experience stable RLOF or CE processes, and a DWD is born eventually. \citet{Han1998} also pointed out two extra formation channels for producing DWDs. One is a double spiral-in channel, i.e., both components are in the giant branch (first RGB or AGB) when the CE is formed. The binary containing two degenerate cores and the ejection of the CE may leave a DWD system. It needs a fine-tuning of initial binary parameters; thus, the possibility of this channel is quite small. Another channel is the CE, followed by the tidally enhanced stellar wind process, i.e., the primary has lost all its envelope via stellar wind before it fills its Roche lobe in the red giant branch. This channel is strongly dependent on the strength of the tidally enhanced stellar wind. Recently, \citet{Macleod2020} found that the tidal enhancement of stellar wind is less than predicted by \citet{Tout1988}. If the tidal enhancement of stellar wind is weak enough, this channel will become less important \citep{Han1998}.

\begin{figure*}
    \begin{minipage}[t]{0.5\textwidth}
        \centering
        \includegraphics[width=\textwidth]{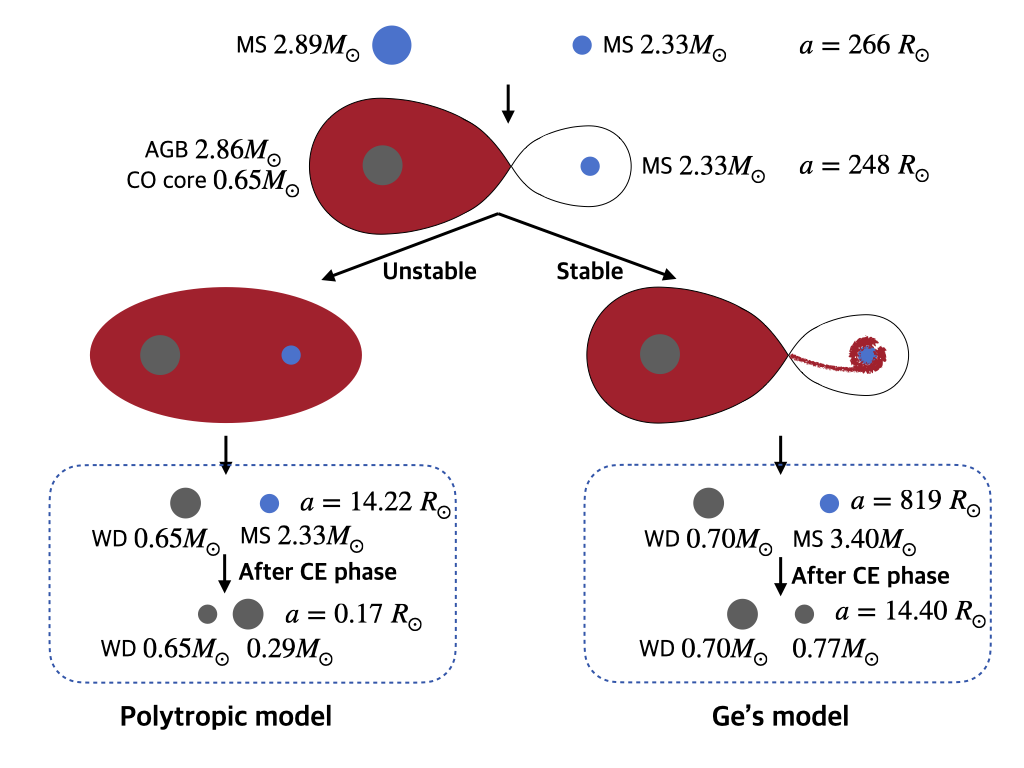}
    \end{minipage}
    \begin{minipage}[t]{0.5\textwidth}
        \centering
        \includegraphics[width=\textwidth]{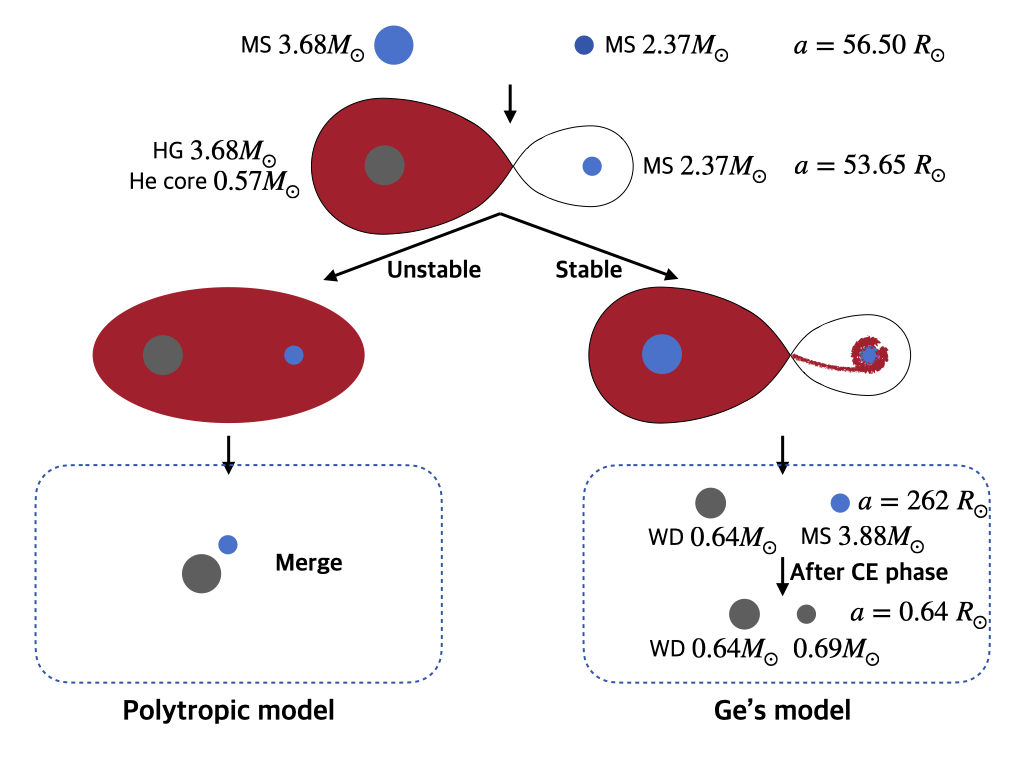}
    \end{minipage}
    \caption{Examples for the formation of DWDs from different mass transfer stability criteria ($\alpha_{\rm CE}=1$). The left panel shows that the DWDs produced from different models have different binary parameters. The right panel suggests that more DWDs would be produced if the first mass transfer phase proceeds in a dynamically stable way. Abbreviations: MS--main sequence, HG--Hertzsprung gap, AGB--asymptotic giant branch, WD--white dwarf. Adapted from \citet{lizw2023}, {reproduced by permission \textcopyright\ ESO}.}
    \label{fig:5}
\end{figure*}

The mass transfer stability is a fiducial question to determine the binary evolution products, as introduced in Section~\ref{sec:3.2}. Recently, \citet{lizw2023} performed detailed BPS simulations and investigated the influence of mass transfer stability criteria on the DWD populations. The examples of such effects are shown in Figure~\ref{fig:5}. It is clear whether the first mass transfer phase is stable directly affects the final DWD properties (left panel). Besides, by adopting the new mass transfer stability criterion (Ge model) of \citet{Geh2020a}, more DWDs would be produced because certain CE mergers are avoided (right panel). Compared to the observational DWDs, they find that the Ge model predicts the merger rate distribution and space density of DWDs, supporting the observations better. Not only DWD populations but other compact object populations should also be significantly affected by the mass transfer stability criteria (e.g., BBH; \citealt{Briel2023,VanSon2022}).

\subsection{NS+WD}
\label{sec:4.2}

There are abundant NS+WD binaries in the Galaxy, based on both the observation and theory \citep{Korol2023}. Binaries of NS+WD in close orbits are supposed to be one type of the progenitor system of UCXB, where the NS accrete material from the degenerate companions (see Section~\ref{sec:5.3} for more details). The NS+WD mergers may have various explosive outcomes, such as Thorne-Zytkow-like objects and long gamma-ray burst \citep{Thorne1977,Paschalidis2011,King2007}. More than 200 NS+WD binaries have been found in the observations, and most of them have low-mass He WD companions (according to the ATNF Pulsar Catalogue \citealt{Manchester2005} in 2024 January). An interesting aspect is that most NSs in this type of binary are found with high spin frequency, e.g., several to several tens milliseconds (known as MSP). It is suggested that the NS should be recycled during the formation process \citep{Tauris2011b}.

\begin{figure}
	\centering
	\includegraphics[width=0.5\textwidth]{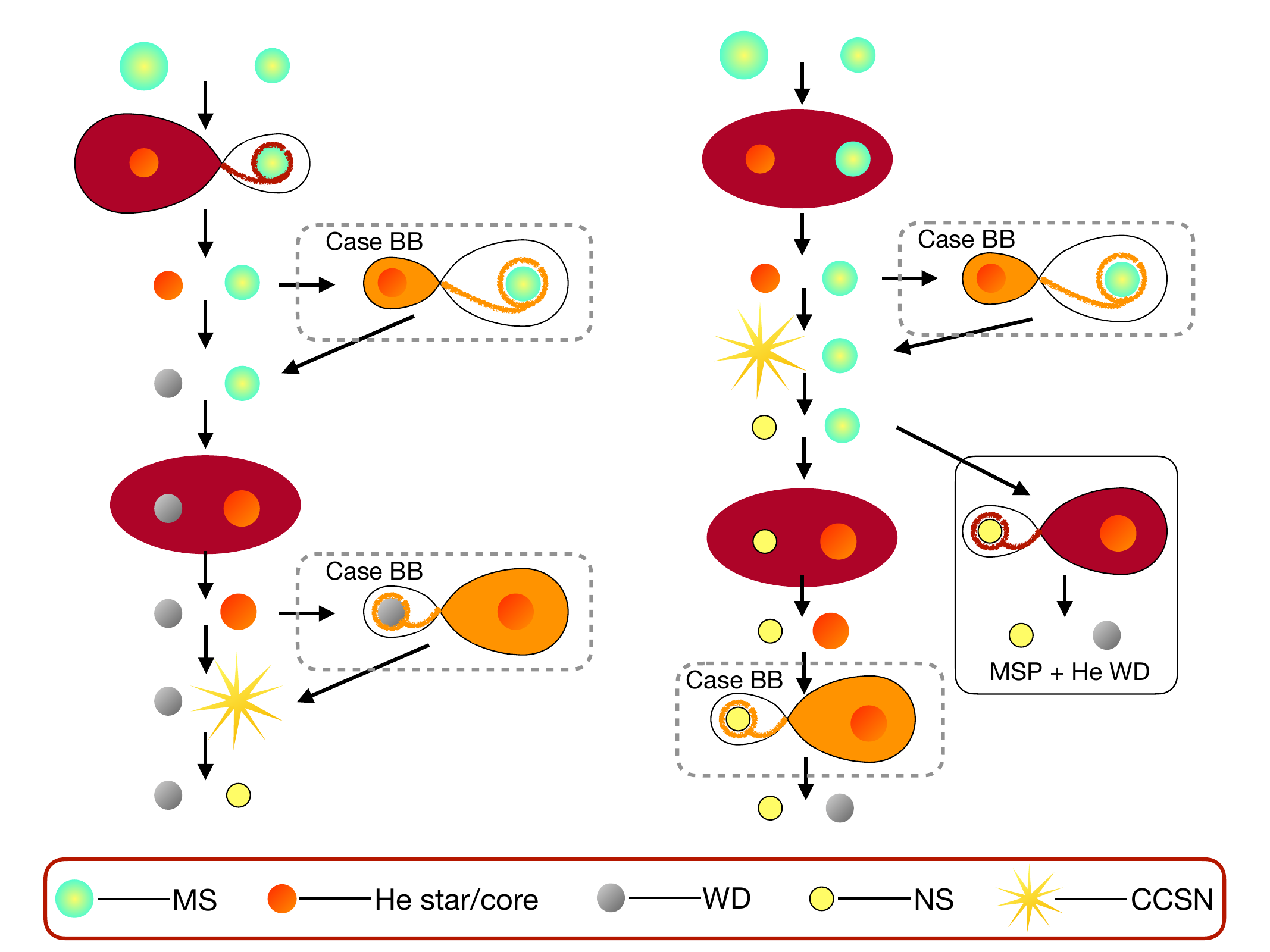}
	\caption{Formation scenarios of NS+WD binaries in close orbits. The Case BB phase may happen when He star has a mass less than $\sim 3.2M_\odot$ (depending on the orbital periods), as shown in the dashed boxes. The MSP+He WD can be formed via stable RLOF with a close orbit when the secondary star (low-mass star) fills its Roche lobe at the MS or early HG stage (depending on the angular momentum loss), as shown in the black solid box. Abbreviations: MS--main sequence, WD--white dwarf, NS--neutron star, MSP-- millisecond pulsar, CCSN--core collapsed supernova.}
    \label{fig:6}%
\end{figure}

Following \citet{Tauris2011a} and \citet{Toonen2018}, we summarised the main formation channel of NS-WD binaries, as shown in Figure~\ref{fig:6}. In the first channel, the progenitor of WD is the initially more massive one. It means that the binary experiences the mass ratio reversal process based on the fact that the progenitor of NS should be more massive. Then, the first mass transfer phase is stable as the massive one fills its Roche lobe. At the termination of mass transfer, it leaves a He star and a massive companion star. The Case BB mass transfer may happen as the He star expands, depending on the binary separation (e.g., \citealt{Habets1986,Tauris2013,Tauris2015,Tauris2017}). Then, the companion star evolves and fills its Roche lobe. The subsequent mass transfer is more likely to be unstable due to the large mass ratio. The binary enters into the CE phase and leaves a more massive He star and a WD companion (or WD precursor) if the CE can be ejected successfully. It is noted that the firstborn He star may not become a WD in this stage because the secondary evolves faster due to the substantial accretion in the first mass transfer phase \citep{Toonen2018}. Similar to the above, the second Case BB mass transfer may happen for the massive He star. After the Case BB mass transfer, the He star is still massive enough to explode as a SN. NS born in this channel does not experience the recycled process and would be observed as a young pulsar. The binary system is generally in eccentric orbits due to the natal kick of the SN. Observationally, some pulsars with WD companions may produced from this channel, such as PSR B2003+46 \citep{Kerkwijk1999}, PSR J1141-6545 \citep{Manchester2000}, and PSR B1820-11 \citep{Lyne1989,Hobbs2004}, and PSR J1755-2550 \citep{Ng2019}.

In the second channel, the progenitor of NS is the initially massive one with mass $\gtrsim 8M_\odot$ (most are in the range of $\sim 11-19M_\odot$ as calculated in \citealt{Toonen2018}). The initial mass ratio is required to be large enough to motivate the CE process in the first mass transfer. Subsequently, the remaining He star from the primary core may expand, and the Case BB mass transfer happens. The NS is born after the SN explosion. The secondary then evolves and fills its Roche lobe. If the mass ratio between the secondary and the NS is large, the mass transfer is more likely to be unstable. After the CE ejection, the binary produces a CO WD/ONe WD and an NS. It is noted that the cass BB phase may happen between the WD precursor (He star) and NS, and the NS is mildly recycled during the accretion process. On the other hand, if the secondary is a low-mass star, i.e., $\lesssim 2.3M_\odot$, the secondary star transfers material to the NS via the stable RLOF. The binary in this stage is observed as LMXB, and the NS is generally recycled as millisecond pulsar \citep{Tauris1999,Tauris2000,Podsiadlowski2002,Neleson2004,Linj2011}. The hydrogen-rich envelope is stripped and leaves a He WD finally. In this channel, the binary orbit is eccentric after the SN explosion, but the subsequent CE phase and the recycled process would circularise the orbit. Many NS-WD binaries in the observation are identified as producing from this channel \citep{Tauris2011a,Tauris2011b}. A distinguished channel for NS binaries is that NSs are born from AIC events (e.g., \citealt{Wangb2022}). However, there is no reported direct detection of AIC events so far, which shows significant uncertainties in this channel (for a recent review, see \citealt{Wangb2020}).

\subsection{NS+NS and BH+NS}
\label{sec:4.3}

\begin{figure}
	\centering 
	\includegraphics[width=0.5\textwidth]{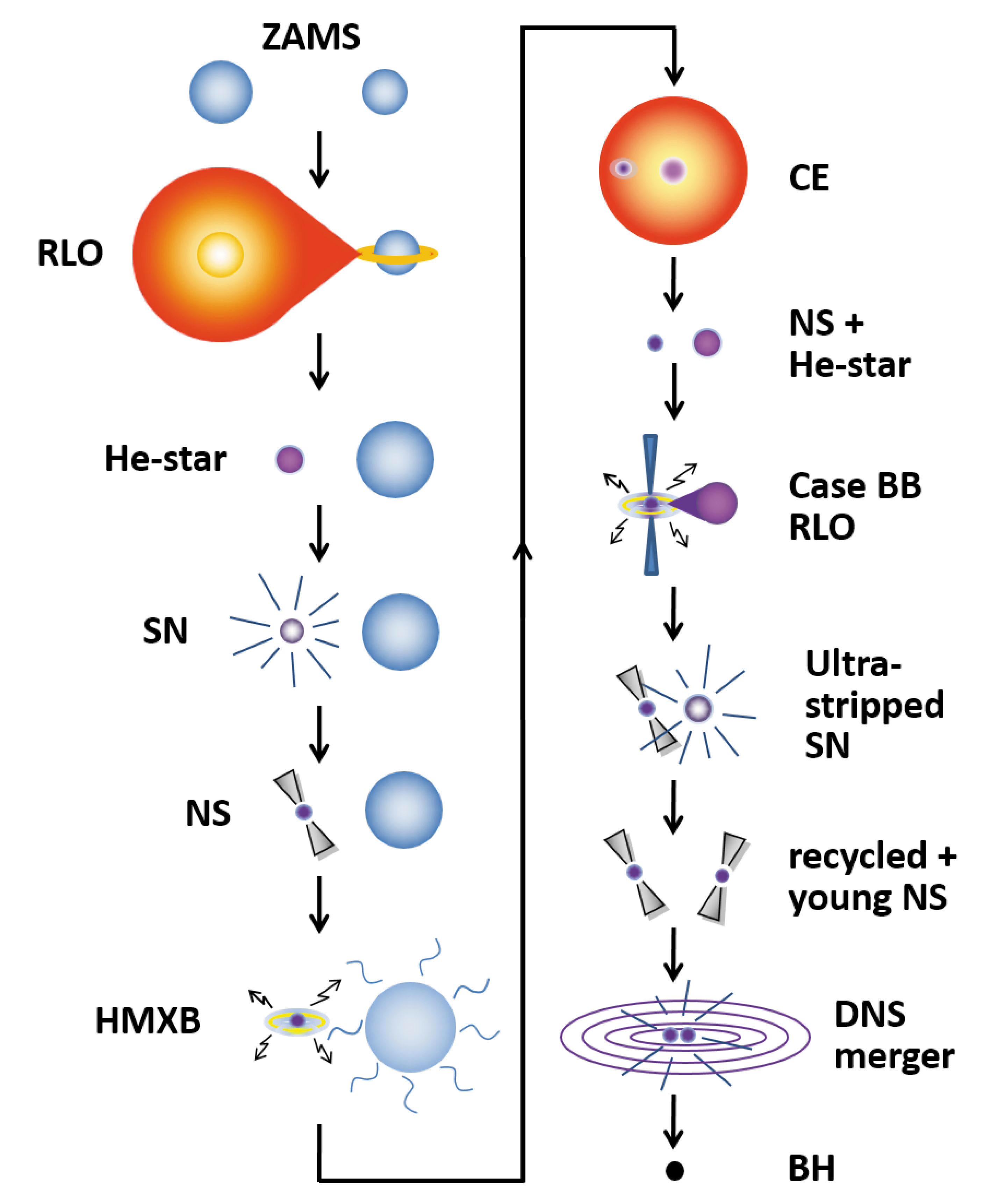}	
	\caption{Illustration of the formation of a DNS system that merges within a Hubble time and produces a single BH. Abbreviations: ZAMS--zero-age main sequence, RLO--Roche lobe overflow, SN--supernova, NS--Neutron star, HMXB--high mass X-ray binary, CE--common envelope, BH--black hole. Adated from \citet{Tauris2017}, {reproduced by permission \textcopyright\ AAS}.}
	\label{fig:7}%
\end{figure}

DNSs are rare events in the Galaxy and are of particular interest since they are generally observable as radio pulsars \citep{Taylor1993,Lyne2004,Lorimer2008,Manchester2013,Desvignes2016,EPTA2023}. Since the first DNS system, PSR B1913+16\citep{Hulse1974,Hulse1975}, found in 1974, only about $\sim 20$ DNS systems are discovered in the radio observations \citep{Manchester2005,Farrow2019}. A few are located in the globular cluster, which is more likely formed from secondary exchange encounters \citep{Fragione2018,Kremer2020}. DNSs found in the Galaxy disk are deemed to be formed from isolated binary evolutions. Many works have been done to construct the evolutionary history of DNSs (e.g., \citealt{Bisnovatyi1974,Wheeler1974,Flannery1975,Srinivasan1982,Heuvel1994,Ivanova2003,Dewi2003,Podsiadlowski2004,Heuvel2004,Dewi2005,Tauris2006}). 

The standard scenario of DNS formation is summarised in Figure~\ref{fig:7}. The initial binary contains a pair of massive OB stars, whereas the primary (initial massive one) is massive enough to end its life in a CCSN. The primary star fills its Roche lobe and strips the hydrogen-rich envelope via the stable RLOF. Meanwhile, the secondary star can grow its mass during the accretion process. If the produced He star has a mass larger than $\sim 1.8M_\odot$ (depending on the wind mass-loss rate of He star), the NS can be born either through the ECSN or the CCSN \citep{Tauris2015,Tauris2017,Woosley2019}. Whether or not the binary survives the following SN explosion depends on the natal kick imparted onto the NS (See Section~\ref{sec:2.3}). The survived binary then becomes observable as a high-mass X-ray binary (HMXB; e.g., \citealt{Kretschmar2019}), where the X-ray is powered by the capture of matter from the stellar wind (\citealt{Martinez2017}; or through RLOF in some cases; \citealt{Chaty2011}).

Once the secondary star fills its Roche lobe, the mass transfer would soon become dynamically unstable, leading to the formation of CE \citep{Paczynski1976}. The binary containing an NS and a He star can be formed if the CE can be ejected successfully. The He star would expand to a large radius with mass $\lesssim 3.2M_\odot$. Therefore, an additional mass transfer phase may be initiated (the so-called Case BB phase; \citealt{Habets1986,Tauris2015}). During the Case BB mass transfer, the He shell is entirely stripped so that the produced NS receives a low kick velocity following the SN explosion (so-called ultrastripped SN; \citealt{Tauris2013,Suwa2015,Tauris2015,Moriya2017,Muller2018}). Besides, the first formed NS can accrete material during the Case BB phase and be recycled to a high spin rate \citep{Tauris2011a,Tauris2017}. If the binary is close enough, the DNS binary will eventually merge due to the GW radiation. 

For the case of the BH+NS binary, the formation scenario is somewhat similar to that of the DNS binary, where the primary star should be more massive to produce a collapsed BH \citep{Jiang2023,Kruckow2018,Chattopadhyay2021,Xing2023}. In some rare cases, the primary star produces an NS, and the secondary makes a BH. It happens when the secondary grows its mass via accretion from the primary star \citep{Kruckow2018}. Analogy to the DWD formation, BH+NS and NS+NS binaries should also experience at least one CE phase. Besides, the kicks imparted onto the NS and BH play an essential role in determining the properties of binary products \citep{Wong2012,Janka2017,Gandhi2019,Zhaoy2023,Kimball2023,Dashwood2023}. Moreover, the complicated SN explosion mechanisms lead to poorly constrained NS and BH masses. Therefore, the theoretical predictions of double compacts containing NSs or BHs are highly uncertain owing to our poor knowledge of these complex physical processes \citep{Tauris2017,Kruckow2018,Breivik2020a,Mapelli2021}.

\subsection{BBH}
\label{sec:4.4}

The formation of BBH is more complicated due to our poor understanding of extremely massive star ($\gtrsim 25M_\odot$) evolution \citep{Marchant2023}. In the traditional formation channel (standard model), BBH is proposed to be produced from HMXB \citep{Belczynski2016}. The evolution route to HMXB is quite similar to that in Figure~\ref{fig:7}, where the firstborn He star is massive enough to collapse as a BH. For HMXB in a wide orbit, the massive companion may fill its Roche lobe as a red supergiant star with a deep convective envelope; the binary then enters into the CE phase \citep{Stevenson2017,Kruckow2018}. The successful CE ejection would lead to the birth of a BH and a massive He star companion, and a close BBH is formed after the collapse of the He star. This channel can successfully explain some massive BBHs with a mass ratio around $1$ in the LIGO observations, such as GW150914 \citep{Belczynski2016}. However, due to the poor understanding of the CE ejection processes, different CE phase assumptions may affect the BBH merger rate prediction by several orders of magnitude \citep{Kruckow2018}. Recently, \citet{Marchant2021} adopted the detailed binary evolution calculation and found that the parameter spaces leading to the successful CE ejection for BBHs as GW sources are significantly lower than previously believed (see also \citealt{Klencki2021}). On the other hand, the HMXB may avoid entering into the CE phase if the massive companion has a radiative envelope, such as the observed X-ray binary SS433 \citep{King1999,King2000,Hillwig2008,Heuvel2017}. The stable RLOF phase can result in a large orbital shrinkage due to the nearly fully non-conservative mass transfer for the BH binaries \citep{Heuvel2017,Wei2023}. Subsequently, some works certified that the stable RLOF in the HMXB phase can make a significant contribution for merging BBHs \citep{Neijssel2019}. The BHs born with experiencing CE and stable RLOF show distinct physical characteristics, e.g., effective spins, which is important to trace the evolutionary scenarios of BBH mergers \citep{Qiny2018,Shaoy2021,Belczynski2020,Zevin2021a,Bavera2020,Callister2021,Marchant2023}. 
 
\begin{figure}[h!]
	\centering 
	\includegraphics[width=\columnwidth]{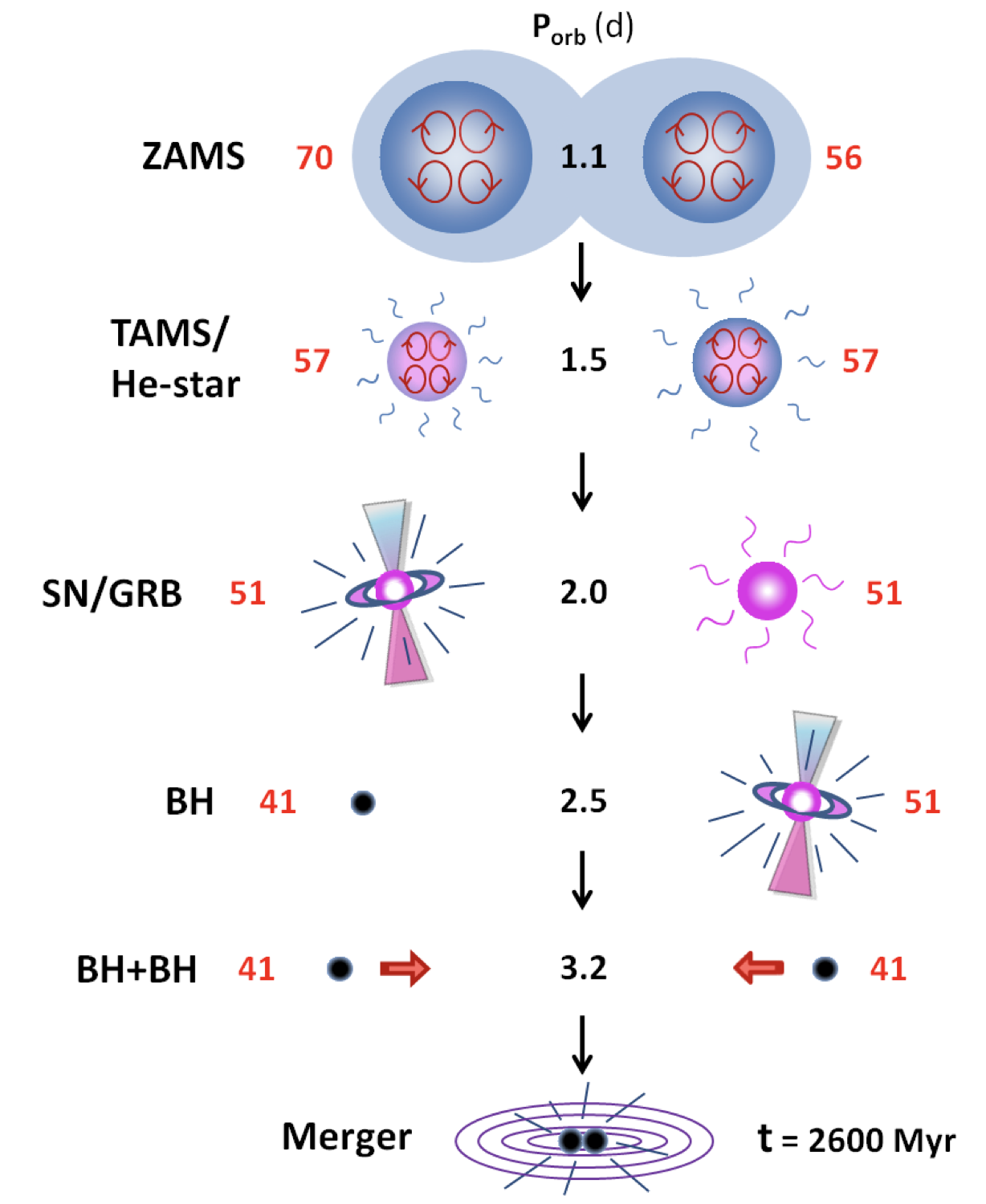}	
	\caption{Illustration of the binary stellar evolution leading to a BH+BH merger with a high chirp mass. The initial metallicity is $Z_\odot/50$, the masses of the stars in solar masses are indicated with red numbers, and the orbital periods in days are given as black numbers. A phase of contact near the ZAMS causes mass exchange. Abbreviations: ZAMS--zero-age main sequence, TAMS--termination age main sequence, He star--helium star, SN--supernova, GRB--gamma-ray burst, BH--black hole. Adated from \citet{Marchant2016}, {reproduced by permission \textcopyright\ ESO}.} 
	\label{fig:8}%
\end{figure}

CHE is also essential for forming BBH in close orbit. In this scenario, the effective rotational mixing induced by the rapid rotation of a star prevents the occurrence of a chemical composition gradient, i.e., the centrally produced He is mixed into the stellar envelope, and the star evolves like a pure He star. Instead of a significant radial expansion, chemically homogenous stars can stay compact and become more luminous \citep{deMink2016,Mandel2016,Marchant2016}. Therefore, this scenario may explain BH born massive. The standard scenario for merging BBH produced from CHE is shown in Figure~\ref{fig:8}. The initial massive binary should be close enough (with an orbital period in the order of $1\; \rm d$) so that both stars evolve with rapid rotation via the tidal interaction. Moreover, the stellar wind may carry away the spin angular momentum. Thus, the CHE only works at low metallicity \citep{Yoon2005,Woosley2006,Brott2011,Kohler2015,Szecsi2015}. The massive binaries evolve into contact where both stars overfill their Roche lobes, leading to the over-contact phase. \citet{Marchant2016} suggested that the over-contact phase can be maintained with no material overflowing the outer Lagrangian point. After the exhaustion of all hydrogen material, the binary encompasses two He stars in a detached stage. Both effects of metallicity and small radial expansion result in a moderate part of material lost from the binary system, leading to massive BHs finally. Due to the existence of an over-contact phase, the binary components can transfer material with each other. Therefore, the two stars should have similar masses after the contact stage, and also, the two BHs produced from this scenario have nearly equal mass. The detailed simulations, including the internal circulation induced by the companion star, suggest that the CHE channel can produce BBHs with mass ratios as low as $0.7-0.8$ \citep{Hastings2020}. The prediction merger rate of BBHs from the CHE channel has been calculated in many works, and the BBHs produced from CHE may dominate the GW detections of BBH mergers from isolated binary evolution (e.g. \citealt{deMink2016,Mandel2016,Marchant2016,Riley2021}). However, one should be careful about the uncertainties in the CHE model, such as the mixing efficiency and the modeling challenge of the over-contact phase, which may bring some uncertainties in the predictions (e.g. \citealt{Songh2016,Marchant2021}). 

A particular isolated binary channel for GW mergers arises from the first generation stars (population III stars) in a theoretical possibility (e.g., \citealt{Kinugawa2014,Inayoshi2017,Tanikawa2021a,Santoliquido2023}). Population III stars are presumed to be formed with enormous masses (e.g. $\gtrsim 100M_\odot$; \citealt{Omukai1998,Abel2002,Bromm2002,Bromm2004,Yoshida2008,Hosokawa2016}), the death of massive population III stars would lead to the formation of massive stellar BHs due to the extremely weak or non-existence of wind mass-loss of metal-free stars \citep{Marigo2001}. Currently, the studies on Population III stars are mainly limited to theoretically numerical simulations; the poor constraints of initial mass function and the star formation history of Population III stars make the predictions from GW mergers somewhat uncertain \citep{Klessen2023}. 

In addition to isolated binary evolution channels for BBH mergers, the dynamical interactions in dense star clusters, e.g., globular clusters, nuclear clusters near the center of galaxies, or active galactic nuclei (AGN), are proven to produce BBH GW sources efficiently \citep{Rodriguez2016,Antonini2016,Mapelli2016,Askar2017,Bartos2017,Stone2017,Banerjee2017,Rodriguez2018,Carlo2019,Fragione2021a,Gerosa2021}. For BBHs produced from dynamical encounters, one distinct characteristic is that the spin and orbital angular momenta should be distributed randomly \citep{Rodriguez2018}. It is possible to detect such signals through the amplitude modulations in the gravitational waveform \citep{Vitale2014}. The dynamical encounters can also lead to hierarchical mergers in a dense environment. The unique GW signatures, e.g., large BH masses and high characteristic spinning, make them distinguishable from BBHs resulting from stellar collapse \citep{Gerosa2021}. There are some other possibilities that may contribute to the GW mergers, such as hierarchical triples, where the so-called Lidov-Kozai cycles would result in significant GW radiation and expedite the inspiral time of binary \citep{Kimball2020,Gerosa2021,Kimball2021}, and primordial BBHs born at the early stage of the Universe \citep{Bird2016,Sasaki2018,Clesse2020}.

\subsection{Other detached binaries}
\label{sec:4.5}

In addition to the double degenerate objects introduced above, there is one type of GW source with only one compact object, i.e., hot subdwarf B (sdB) + compact star. SdB star is He burning star with a thin hydrogen envelope \citep{Heber2009,Heber2016}. Since the hydrogen envelope of sdB has been almost entirely stripped, the sdB radius could be very small. The recently reported sdB binary, TMTS J052610.43+593445.1, has an extremely short orbital period of $20.5\;\rm min$, where the radius of the sdB star is only $0.066\;\rm R_\odot$, about seven Earth radii \citep{Linj2023}. This binary serves as a crucial verification source for future space-borne GW detectors. There are two other sdB binaries in tight orbits as LISA verification sources, i.e., CD-$30^\circ 11223$ of $70.5\;\rm min$ orbit \citep{Geier2013} and HD265435 of $99.1\;\rm min$ orbit \citep{Pelisoli2021}. These three sdB binaries have WD companions, and sdBs with NS and BH companions have not been found yet (exist in theory; e.g., \citealt{Wuy2018,Wuy2020,Gotberg2020}). The formation scenarios of sdB binaries require similar envelope-stripped processes as that of WD binaries, and sdB binaries in close orbit are unquestionably produced from the successful CE ejection \citep{Han2002,Han2003,Geh2022,Geh2023b}. 

\section{Accreting Binary}
\label{sec:5}
Accreting binaries with compact objects are largely found in the observations. The accretors could be WD, NS, or BH. The orbital periods of accreting compact objects occupy an extensive range, from several minutes to several hundred days. Many accreting compact objects show strong X-ray, such as super soft X-ray source (SSS; WD accretor), LMXB, and HMXB (NS or BH accretor), etc., which are crucial to understanding the accretion physics. The scope of this review focuses on the GW sources. Accreting binaries with typical orbital periods larger than several hours would not contribute to the GW observations (both of LISA and LIGO), which will not be introduced in detail here. A detailed discussion of all types of accreting compact objects can be found in the recent reviews of \citet{Chaty2022,Belloni2023b}.
\subsection{CVs}
\label{sec:5.1}

Cataclysmic variables (CVs) are accreting WDs (most are CO WDs) with hydrogen-rich donors (low-mass MS stars or sub-stellar objects). The typical CV period ranges from $\sim 75$ min to $\sim 1$ day \citep{Knigge2011,Pala2020,Inight2021}. The nature of short orbital periods for CVs suggests that almost all CVs progenitors have experienced one CE ejection phase. The successful ejection of CE leads to a detached WD+MS binary with orbital periods around $1\;\rm d$. For a low-mass star ($\lesssim 1.5M_\odot$), the surface magnetic field is assumed to arise from the differential rotation between the convective envelope and radiative core. The so-called magnetic braking carries away the system’s angular momentum and shrinks the orbit until the MS star fills its Roche lobe, and a CV is born. One extraordinary characteristic of CVs is the period gap of $2-3\;\rm$ hours found in the observations. In the standard frame of CV evolution, the reason for the period gap is caused by the disappearance of the magnetic field when the donor becomes fully convective with mass $\lesssim 0.3M_\odot$. The system evolves as a detached binary due to the absence of magnetic braking. Then, the orbital shrinkage is dominated by the GW radiation, the donor refills its Roche lobe at an orbital period of about $2\;\rm h$, and the accretion process resumes. The orbital period would evolve to a period minimum when the donor cannot sustain the hydrogen burning and becomes a hydrogen-rich degenerate object \citep{Kalomeni2016,Belloni2023a,Sarkar2023b}, then the orbital period increases as the donor expands in response to the mass loss. The period minimum depends on the compactness of the donor, i.e., the central He abundance \citep{Kalomeni2016,Belloni2023a}. 

\subsection{AM CVn}
\label{sec:5.2}

AM CVn binaries are ultra-compact interacting WD binaries with typical orbital periods in the range of $\sim 5-65\;\rm min$ \citep{Ramsay2018}. The closer orbits than CVs suggest that the donor should be more compact. Most AM CVn stars do not show hydrogen in their spectra, suggesting its origin from He-rich donor \citep{Solheim2010}. Three main formation channels lead to the birth of AM CVn binaries, i.e. (1) He WD channel: a detached DWD with close orbit is formed first, and the low-mass WD fills its Roche lobe due to the orbital shrinkage caused by the GW radiation \citep{Kremer2017,Chenh2022b}. (2) He star channel: similar to the He WD channel, but the AM CVn progenitor is a detached WD and He star binary if the He star fills its Roche lobe before being a WD (means the timescale of orbital shrinkage due to the GW radiation is shorter than the lifetime of a He star), an AM CVn binary is formed \citep{Sarkar2023a}. (3) Evolved CV channel: as introduced in Section~\ref{sec:5.1}, the period minimum of a CV binary is determined by the compactness of the donor. Then, if the donor of a CV is a slightly evolved MS star, the loss of hydrogen-rich envelope finally leads to a degenerate he-rich core, and the orbital period of a CV could evolve well below $\sim 60\;\rm min$ \citep{Sarkar2023b,Belloni2023a}. 

\textbf{WD channel:} In the early studies, the WD channel is presumed to be the main formation channel of AM CVn (e.g. \citealt{Tutukov1996,Nelemans2004}). A critical issue in the WD channel is the mass transfer stability between two WDs \citep{Marsh2004,Motl2007,Dan2012}. In addition to the radial response of the WD donor, the tidal coupling should also play a vital role in the case of close DWD. As shown in \citet{Marsh2004}, the WD channel significantly contributes to the AM CVn stars only when the synchronizing timescale acts on a time scale substantially less than $1000\;\rm yr$. \citet{Shen2015} examined the stability of DWD mass transfer with considering the sturcture of WD accretor. The author found that the initially transferred hydrogen-rich material would lead to the nova outburst, and the dynamical friction within the expanding nova shell shrinks the orbit and causes dynamically unstable mass transfer. Finally, the interacting DWD may merge. This conclusion rudely excludes the WD channel of AM CVn stars. However, the nova outburst is a rapid process, and the interaction between the nova shell and the companions should be considered in a hydrodynamical way. Moreover, the recent observations of Roche lobe filling hot subdwarf binaries \citep{Kupfer2020a,Kupfer2020b,Lij2022} obviously conflict with this scenario. As calculated by \citet{Bauer2021}, several times of shell flashes would occur when the hot subdwarf transfers hydrogen-rich material onto the WD surface. 

The observed sample in the ELM Survey may provide some hints for the stability problem of interacting DWDs. \citet{Brown2020} analyzed the merger rate of the DWDs with ELM WD companions in the complete sample of ELM Survey and found that the merger rate of ELM WD binaries is about 25 times larger than the birthrate of AM CVn stars (See also \citealt{Brown2016}). This finding suggests that most ELM WD binaries will merge instead of forming AM CVn binaries. \citet{Kilic2016} assumed that only the observational sample of ELM Survey experiencing disc accretion could form AM CVn binary via stable mass transfer and found the formation rate of stable mass transfer systems from binary ELM WDs is coincident with the AM CVn formation rate found in the observations \citep{Carter2013}. The physics behind is still unclear; possibly, the synchronizing torque on the DWDs should act on a time-scale much larger than $1000\;\rm yr$ \citep{Marsh2004}. Above all, the stability of interacting DWDs remains an open question and deserves further investigation.

\textbf{He star channel: }In the He star channel, the mass transfer begins before the exhaustion of He burning. Many works on this channel assumed a pure He star with no hydrogen envelope (e.g., \citealt{Tutukov1996,Nelemans2001b,Postnov2014,Liuw2022}). However, He star is supposed to possess a thin hydrogen envelope at its birth, which is classified as a hot subdwarf star in the observations. The parameter space leading to AM CVn binaries for He stars with varied hydrogen envelopes is different \citep{Bauer2021}. Therefore, it deserves a further study of the He star channel with a detailed model. On the other hand, when the AM CVn evolves to the period minimum, the He star would become a degenerate object like He WD \citep{Wangb2021}, encountering a similar problem of the mass transfer stability between two WDs as introduced above.  

\textbf{Evolved CVs channel: }The evolved CVs channel has been considered insignificant for forming AM CVn previously \citep{Nelemans2004}. On the one hand, the range of parameter space of initial CV progenitors leading to AM CVn is very narrow \citep{Goliasch2015,Kalomeni2016}. The fine-tuning parameter space then predicts a low AM CVn rate. On the other hand, most AM CVn binaries show hydrogen-deficient spectra, which are distinct from those of typical CVs. Recently, the evolved CV channel has attracted more and more attention (e.g., \citealt{Liuw2021,Belloni2023a,Sarkar2023a}) with the increase of observational sample of CVs \citep{Pala2020,Green2020,ElBadry2021,Pala2022}. \citet{Belloni2023a} revisited the evolved CV channel by adopting the newly proposed magnetic braking description of the CARB model \citep{Van2019b}. In the CARB model, the magnetic braking would carry away more orbital angular momentum than that of classical ``Skumanich'' MB laws \citep{Skumanich1972}. As verified by \citet{Belloni2023a}, no fine-tuning of the initial orbital parameter is needed to produce AM CVn with the CARB magnetic braking prescription. Meanwhile, the accretion from the developed He core, which has developed prior to the onset of mass transfer, can naturally explain the hydrogen-deficient spectra of AM CVn stars. 

All three of these channels are suitable for producing AM CVn binaries. However, we need to find out which channel dominates. A systemic investigation combining all possible channels for AM CVn stars in the population synthesis is urgently required. 

\subsection{UCXB}
\label{sec:5.3}

UCXB is a subtype of low-mass X-ray binaries with hydrogen-deficient donors. UCXBs have similar orbital periods as AM CVn stars, and the main difference is that the UCXB possesses a massive accreting compact object of either an NS or a BH \citep{Belczynski2004}. So far, Several UCXBs (or candidates) have been found, including persistently active X-ray sources and transient objects \citep{Armas2023}. Many efforts have been devoted to explaining the observational characteristics of UCXBs. Nevertheless, it still needs to be solved to elucidate the evolutionary channels (e.g., \citealt{Nelemans2010a,Nelemans2010b,Heinko2013}). Similar to AM CVn binary, the donor in a UCXB could be a He WD, a non-degenerate (or semi-degenerate) He star, or an evolved MS star \citep{Nelemans2010a,Zhuc2012,Haaften2013,Lv2017,Sengar2017,Shaoy2020,Chenh2021,Wangb2021,Guoy2022,Qink2023a,Qink2023b}. Comparing with AM CVn stars of WD accretors, the compact accretors in UCXB can form via CCSN or AIC. In the latter scenario, the NS accretor is produced from a massive ONe WD accumulating material until reaching the Chandrasekhar mass limit \citep{Liud2023}. Similarly, the BH accretor can be formed from a massive NS close to the maximum mass of an NS via the AIC process \citep{Chenh2023}. 

An important issue concerning He WD channel of UCXBs is the critical WD mass of mass transfer instability, which directly determines the number of UCXBs in the Galaxy \citep{Nelemans2010a}. However, the long-term mass transfer stability remains a problem where the numerical computations give distinct results. Several works based on the semi-analytic methods found that the critical WD mass is approximately $0.37-1.25M_\odot$ \citep{Verbunt1988,Paschalidis2009,Yungelson2002,Haaften2012,Yus2021}. The recent simulations of \citet{Chenh2022a} with considering the full structure suggested that all He WDs with masses ranging from $0.17-0.45M_\odot$ would undergo stable mass transfer. Moreover, \citet{Bobrick2017} carried out hydrodynamic simulations and found that disk winds are very efficient at removing angular momentum so that the critical WD mass is only about $0.2M_\odot$, which is much lower than that of hydrostatic simulations. The difference depends on the details of model inputs, which arises from the poor understanding of the accretion physics with compact objects. The future GW observations of UCXBs will improve our knowledge about this issue (See Section~\ref{sec:6}).

\section{Compact binaries as GW sources}
\label{sec:6}

\subsection{high-frequency GW sources}
\label{sec:6.1}
GW detectors in the ground aim to detect high-frequency GW sources, i.e., sources with frequencies larger than $\sim 1\;\rm Hz$. Several types of sources would contribute to this frequency domain, such as compact binary coalescences, short-duration GW bursts, non-axisymmetric spinning neutron stars, and the stochastic GW background \citep{Abbott2009,Abbott2022}. Here, we mainly focus on the first one, i.e., binary coalescence with two BHs, two NSs, or BH+NS. Given that WD has a typical radius of several thousand kilometers, there is no contribution of compact binaries with WD companions for ground-based GW detectors (but there may be high-frequency GW signals if the binary contains a non-axisymmetric spinning NS; e.g., see \citealt{Abbott2022}). 

Up to now, nearly 100 GW transients (or candidates) are reported by the combined LVK Collaboration. Most GW events are BBH mergers ($\gtrsim 80$), coupled with some NS+NS mergers and mixed BH+NS mergers \citep{Abbott2023a,Abbott2023b}. The typical BH masses in the GW catalog are larger than $20M_\odot$ while those from EM observations are typically less than $20M_\odot$. It is not surprising since GW detectors are inclined to find sources with massive objects from which powerful GW signals can be emitted. The massive BHs are supposed to be formed in a low-metallicity environment, where the stellar winds can be largely depressed (but see \citealt{Bavera2023} who found stellar at solar metallicity can produce BHs with masses beyond $30M_\odot$). Many works aimed to investigate the origin of these compact binaries, and most of them can be well explained by the isolated binary evolution (e.g. \citealt{Mandel2016,Belczynski2016,deMink2016,Marchant2016,Kruckow2018,Spera2019}). Nevertheless, other formation pathways without involving binary interaction cannot be excluded \citep{Mandel2022}. Besides, there are many exotic compact objects in the catalog, of which the formation scenarios are still in debate, such as the highly massive BBHs with BH components inside the PISN mass gap (e.g. GW190521; \citealt{Abbott2020a,Abbott2020b}) and the compact binaries with extreme mass ratio (e.g. GW190814 and GW200210; \citealt{Abbott2020c,Abbott2023b}). 

The measured merger rate densities for DNS, BH+NS and BBH mergers are $10-1700\;\rm Gpc^{-3}\;\rm yr^{-1}$, $7.8-140\;\rm Gpc^{-3}\;\rm yr^{-1}$ and $16-61\;\rm Gpc^{-3}\;\rm yr^{-1}$ , respectively (local values at $z=0$; \citealt{Abbott2023c}). These inferred values depend on the selection effects and have large uncertainties \citep{Mandel2022}. With the improvement of the detector sensitivity in the third-generation GW instruments, the merger rates in the observations can be well constrained \citep{Kalogera2019,Ng2021}. A large number of theoretical works have been performed in calculating the merger rate densities, but the predicted values for individual channels may vary by two or three orders of magnitude, which arises from the lack of understanding of the evolutionary phases of massive binaries (e.g., \citealt{Shaughnessy2010,Mennekens2014,deMink2015,Dominik2015,Mapelli2017,Ablimit2018,Chruslinska2018,Giacobbo2018,Klencki2018,Mapelli2018,Kruckow2018,Artale2019,Baibhav2019,Chruslinska2019,Eldridge2019,Neijssel2019,Spera2019,Belczynski2020,Giacobbo2020,Santoliquido2020,Santoliquido2021,Shaoy2021,Mandel2016,Marchant2016,Buisson2020,Riley2021}). Although many scenarios have been proposed to explain the observed GW populations, the relative contribution is still being determined \citep{Zevin2021b,Zevin2021a}. With the remarkable characteristics (e.g., mass, spin, eccentricity) of some GW events, one could distinguish the individual channels for these mergers (e.g., \citealt{Qiny2018,Qiny2019,Bavera2020,Zevin2021a,Qiny2022,Qiny2023}). 
Due to the limited sensitivity, current GW detectors can only be capable of observing mergers in the local Universe ($z \lesssim 1$; \citealt{Abbott2019a,Abbott2019b,Abbott2020d,Roulet2020,Venumadhav2020,Abbott2023a,Abbott2023b}). The era of third-generation GW detectors, e.g., the Einstein Telescope \citep{Punturo2010} and Cosmic Explorer \citep{Reitze2019}, would enlarge the merger sample by a factor of $100000$, the unbiased (or less-biased) observational sample of GW merger make sure to improve our understanding of the nature of GW mergers. 

There is the theoretical possibility that GW mergers arise from the first generation stars (population III stars; e.g., \citealt{Kinugawa2014,Tanikawa2021a,Santoliquido2023}). Population III stars are supposed to produce massive BHs due to the extremely weak or non-existence of wind mass-loss \citep{Marigo2001}. Based on isolated binary evolution channels, \citet{Santoliquido2023} found the BBH merger rates of population III stars at $z=0$ are typically less than $0.2\;\rm Gpc^{-3}\;yr^{-1}$, which are lower than the local BBH merge rate density inferred from LVK data by about two orders of magnitude (see also \citealt{Kinugawa2014,Kinugawa2016,Belczynski2017,Tanikawa2021b}). However, at high redshift ($z\gtrsim 8$), the BBH merger rates from population III stars increase spanning from $\sim 2$ to $\sim 30$$\;\rm Gpc^{-3}\;yr^{-1}$. These merger events are expected to be captured by the next-generation ground-based GW interferometers with horizons up to $z\gtrsim 100$ \citep{Ng2021,Ng2022a}, which possess the ability to probe the merger of BBHs in the early Universe \citep{Singh2022,Ng2022b,Santoliquido2023}. 
 
In addition to isolated binary evolution channels for BBH mergers, the dynamical interactions in dense star clusters, e.g., globular clusters or nuclear clusters near the center of galaxies, are proven to produce BBH GW sources efficiently \citep{Rodriguez2016,Antonini2016,Mapelli2016,Askar2017,Banerjee2017,Bartos2017,Stone2017,Park2017,Rodriguez2018,Fragione2021a,Gerosa2021,Liub2021}. The dynamical formation channel allows a chain of hierarchical mergers and is capable of producing massive BHs in the pair-instability mass gap, such as GW190521 \citep{Fragione2020,Anagnostou2022}. Hierarchical triple systems formed in globular clusters or Galactic field are also likely to produce GW mergers via the Kozal-Lidov resonances \citep{Kimball2020,Gerosa2021,Kimball2021}, but the contribution to LVK observations should be small ($\sim 1\%$; e.g., \citealt{Silsbee2017,Antonini2017,Martinez2020}). Nevertheless, the detectable eccentricities in the LVK frequency band of the hierarchical triple channel may provide important hints to distinguish the formation channels in the GW observations \citep{Rodriguez2018,Martinez2020,Zevin2021b,Zevin2021a}.

\subsection{Low-frequency GW sources}
\label{sec:6.2}

Our Galaxy comprises several hundred billion stars, including $\sim 10^8$ double compact objects. The double compact objects in the Galaxy have been studied in several decades \citep{Webbink1984,Han1995,Evans1987,Han1998,Nelemans2001a,Nelemans2001b,Nelemans2001c,Liuj2009,Liuj2010,Yus2010,Ruiter2010,Nissanke2012,Toonen2012,Korol2017,Korol2019,Lamberts2019,Breivik2020a,Korol2021,lizw2019,lizw2020,lizw2023}. We believe that all types of double compact objects exist in our Galaxy, the expected numbers and those can be detected by LISA are summarised in Table~\ref{tab:1}. It is not surprised that DWD binaries are the most common compact binaries in the Galaxy, mainly on account of more low-mass and intermediate-mass stars based on the initial mass function. The abundant DWD sources in the LISA frequency domain would provide significant information and shed light on the key questions in binary interaction physics. For DWDs with frequency in the range of $\sim 10^{-4}-5\times 10^{-3}\;\rm Hz$, corresponding to orbital period of $5-0.1\;\rm h$, more than one sources in the LISA minimum resolvable frequency bin, the superpositions of GW signals make up the particular noise of the instrument, known as GW foreground noise \citep{Ruiter2010,Nissanke2012}. Unlike the instrument noise, the GW foreground carries essential information about DWD populations, which may offer an opportunity to probe the Galactic structure \citep{Breivik2020b,Georgousi2023}. Sources with higher frequency can be individually resolved by the LISA, also known as resolvable sources. At least several thousand DWDs are resolved by LISA, which could enlarge the observational DWD sample by several ten or hundred times \citep{Yus2010,Korol2017,Lamberts2019,lizw2020}. The number distributions of the well-measured binary periods and chirp masses can limit the binary evolution processes, particularly the CE physics \citep{LISA2023}. 

The second numerous source is the NS+WD systems. Though the expected number is lower than that of DWDs with about two orders of magnitude, NS+WD systems are of great interest due to the existence of eccentricity, which may shed light on the NS natal kicks \citep{Ruiter2019,Breivik2020a,Korol2023}. It is challenging to distinguish the NS+WD system from the DWD systems due to the similar frequency ranges occupied for these two types of binaries (as shown in Figure 4 of \citealt{Korol2023}). Thanks to the natal kick imparted on the NS in the NS+WD binaries, the identification of eccentric source would distinguish NS+WD binaries from DWDs beyond doubt \citep{Moore2023}. On the other hand, for sources with frequencies larger than $\sim 2$ mHz, the frequency changes are able to be detected by LISA, which then gives the chirp mass. The typical chirp masses for DWDs are in the range of $0.2-0.4M_\odot$ with a long tail extending up to $1M_\odot$ \citep{Yus2010,lizw2019,Korol2017}. In comparison, the NS+WD binaries have chirp masses in the range of $0.35M_\odot-1.2M_\odot$ with a peak around $\sim 0.8M_\odot$ \citep{Korol2023}. Therefore, it has a large possibility to identify an NS+WD binary with a large chirp mass (e.g., $\gtrsim 0.8M_\odot$). 

DNSs are one of the most important multimessenger sources in astrophysics. One famous source is the GW170817, the first GW signal of Double NS observed in LIGO and followed by the EM signals of short gamma-ray burst (GRB 170817A) and the associated kilonova \citep{Abbott2017b,Abbott2017c,Abbott2017d,Cowperthwaite2017,Savchenko2017,Troja2017,Smartt2017,Goldstein2017}. LISA is hope to detect the GW signals of DNSs millions of years before their coalescence. The challenges of modeling DNS populations in the Galaxy arise from the poor constraint of the natal kick and the massive binary interaction \citep{Yus2015,Tauris2017,Kruckow2018,Storck2023}. In a recent study, \citet{Wagg2022} performed detailed population synthesis for DNS, BHNS, and BHBH populations in the Galaxy, considering various physical inputs in binary evolutions. The results show that LISA would detect about 10 to several $100$ double compact objects with BH or NS companions within four years, and the numbers doubled for a 10 year survey time, among which about half of sources can be distinguished from DWDs based on their mass or eccentricity and localization. For a given source, the maximum detectable distance for LISA (also ground-based detectors) is scaled with $\mathcal{M}^{5/3}$. Therefore, one could expect that WD and NS binaries are detected by LISA within the local group \citep{Korol2018,Korol2020,Keim2023}. While, BBHs with typical masses of several tens of solar mass may be detected in distant galaxies, e.g., for GW150914-like BBHs located several hundred Mpc away \citep{LISA2023}. 

The predictions of semi-detached binaries as GW sources are somewhat uncertain. CVs are close binaries with WDs accreting material from the low-mass MS companions with typical orbital periods of $75\;\rm min $ to $1\;\rm d$. Due to the large size of the non-degenerate companions, CVs would evolve towards the minimum orbital period near $P_{\rm min}\simeq 75\;\rm min$. Many CVs then contribute to the GW signals in the range of $0.2-0.4\;\rm mHz$, as parts of the foreground noise \citep{Scaringi2023}. The closest CVs may be individually resolvable by the GW detectors. Nevertheless, it is not trivial to pick out CVs from numerous DWDs \citep{Scaringi2023}. AM CVn binaries and UCXBs have shorter orbital periods in the range of $\sim 5-65\;\rm min$. The indeterminate progenitors of AM CVn and UCXBs create difficulties in the predictions of LISA sources. Fortunately, we now know several LISA verification sources of AM CVn and UCXBs in the EM observations \citep{Kupfer2018,Kupfer2023}. For example, HM Cnc with an orbital period of $5.4\;\rm min$ is the reference source for the TianQin project \citep{Luoj2016,Huangs2020}. The adjacent observation of LISA and TianQin would capture the GW signals from these accreting binaries, and the precise measurements of the orbital parameters should make a great improvement in the understanding of the formation scenarios of these binaries \citep{Breivik2018}.

\begin{table}
\begin{tabular}{l c c } 
 \hline
 Source & $N_{\rm Gal}$ & $N_{\rm det}$ \\ 
 \hline
 WD+WD  & $\sim 10^8$   & $6000-30000$   \\
 NS+WD  & $\sim 10^7$   & $100-300$   \\
 NS+NS  & $\sim 10^5$   & $3-35$   \\
 BH+WD  & $\sim 10^6$   & $0-3$   \\
 BH+NS  & $\sim 10^5$   & $2-198$   \\
 BH+BH  & $\sim 10^6$   & $6-154$   \\
 \hline
 \hline
 CVs  & $\sim 10^6-10^7$   & $\mathcal{O}(1)$   \\
 AM CVn  & $\sim 10^7$   & $200-2700$   \\
 UCXB  & $\sim 10^4$   & $\mathcal{O}(100)$   \\
 \hline
\end{tabular}
\caption{The estimated number of compact binaries from isolated binary evolution in the Galaxy. The first column is for the sources, the second is for the present-day number of sources in the Galaxy, and the last is for the estimated number of sources detected by LISA. The estimated numbers are extracted from the binary population synthesis or semi-analytic results, WD+WD from \citet{Nelemans2001c,Yus2010,Nissanke2012,Korol2017,Korol2019,Lamberts2019,lizw2020,lizw2023}, NS+WD from \citet{Toonen2018,Korol2023}, NS+NS, BH+NS and BH+BH from \citet{Wagg2022}, BH+WD from \citet{Nelemans2001c}, CVs from \citet{Goliasch2015,Scaringi2023}, AM CVn from \citet{Nelemans2004,Kremer2017}, and UCXB from \citet{Belczynski2004,Zhuc2012,Chenw2020}.}
\label{tab:1}
\end{table}

\subsection{Multi-messenger GW sources}
\label{sec:6.3}

Multi-messenger observations are of great importance in understanding the nature of GW mergers and the hidden physics. The most remarkable event is the GW170817, which is the only source with the synergistic observations of GW and EM detectors so far \citep{Abbott2017b,Abbott2017c,Abbott2017d,Cowperthwaite2017,Savchenko2017,Troja2017,Smartt2017,Goldstein2017}. GW170817 provides the solid evidence of the connection among DNS merger, short gamma-ray burst and kilonova, and also offers an opportunity to probe the properties of matter at the extreme conditions \citep{Margalit2017,Kasen2017,Annala2018,Pian2017}. Another type of merger may produce an EM counterpart associated with the GW signal at merger known as BH+NS binaries. Several BH-NS merger candidates have been reported in the LVK catalog, e.g., GW190814, GW200105, GW20115 \citep{Abbott2023a,Abbott2023b}. Unfortunately, there is no associated EM counterpart confirmed in the follow-up observations \citep{Goldstein2019,Hosseinzadeh2019,Coughlin2020,Thakur2020,Alexander2021,Anand2021,Kilpatrick2021}). In fact, the EM counterpart for a BH+NS merger is expected only when the NS is tidally disrupted, rather than directly plunging into the BH. The emission of EM signals then requires that BH with high spin and NS with soft equation-of-state \citep{Fragione2021b}. It is a more stringent requirement than that of DNS mergers. But the EM counterparts for BH+NS mergers can still be expected in the future \citep{Zhuj2022,Neill2022,Dorazio2022,Gupta2023,Gompertz2023,Steinle2023}. 

{Many telescopes worldwide are dedicated to detecting the EM counterparts of GW sources, of which the recent projects in China will undoubtedly play an important role. The Space Variable Objects Monitor (SVOM; Chinese-French mission; \citealt{Yus2020}) and Gravitational wave high-energy Electromagnetic Counterpart All-sky Monitor (GECAM; \citealt{Zhangd2019}) missions aim to monitor the GW-associated GRBs. The results will guide the follow-up observations of EM counterparts and verify the corresponding objects. The Einstein Probe (EP; \citealt{Yuanw2022}) is an all-sky survey of transients and explosive objects at the X-ray band and provides a powerful tool to discover GW merger events. The ground-based telescopes like Multi-channel Photometric Survey Telescope (MEPHISTO; \citealt{Yuanx2020}) and Wide Field Survey Telescope (WFST; \citealt{Wangt2023}) will carry out the time-domain survey at the optical band and have the outstanding capability of follow-up observations with high sensitivity after the GW merger events. These large-scale multi-wavelength instruments would provide a unique opportunity for us to understand the merger processes of extremely dense objects.}

BBHs are not expected to be directly discovered in the EM observations. However, the multiband GW observations are available for massive coalescing GW mergers and would deliver important information on the formation scenarios \citep{Sesana2016,Breivik2016,Zhaoy2023,Isoyama2018,Gerosa2019,Vitale2016,Lius2020,Sesana2020,Seto2022,Xuan2023}. For GW150914-like BBHs, the GW frequency is about several mHz ten years prior to the coalescence, well within the LISA band (also other mHz GW detectors; \citealt{Sesana2016}). It is possible to find GW signals from GW150914-like binaries in advance in the LISA data so that we can know precisely when and where they would appear in the ground-based GW detectors. Most importantly, sources emitting low-frequency GW signals may have high eccentricities and can be well constrained in the LISA observations \citep{Nishizawa2016,Seto2016}. One can use the eccentricity to distinguish the formation channels of BBHs \citep{Breivik2016}. 

Except for BBHs, other compact binaries are potentially multi-messenger sources that can be detected in EM observations and space-borne GW detectors. The multi-messenger study will allow us to get more precise information for these compact objects, such as distance, sky location, and individual masses \citep{Shah2014a,Shah2014b,Littenberg2019}. Up to now, we have dozens of LISA verification or detectable sources, including detached DWDs, AM CVn, UCXB, and sdB binaries \citep{Kupfer2018,Burdge2019a,Kupfer2023,Linj2023,Kosakowski2023b}, where the AM CVn binaries and DWDs provides $\sim 90\%$ of the total number. The running and planned instruments, e.g., GAIA, ZTF, and LSST, are hoped to increase the number substantially \citep{Bellm2019,Gaia2016,LSST2009}. \citet{Korol2017} explored the prospects for detecting detached DWDs in GAIA, LSST, and LISA and found that about $\sim 80$ DWDs can be detected through EM and GW radiation. \citet{lizw2020} suggested that most of these multi-messenger DWDs should contain an ELM WD companion since ELM WDs can sustain a high-luminosity phase for a long time and are more easily detected in the EM observations (see also \citealt{Chenx2017,lizw2019}). A large number of DWD detections then provides an opportunity to trace the Milky Way potential and constrain the rotation curve of our Galaxy \citep{Korol2019}. For the accreting binaries, e.g., UCXB and AMCVn, the combination of EM and GW observations would have well-constrained frequency derivative \citep{Breivik2018}, shedding light on the accretion physics with compact objects. Combined EM and GW observations can also solve some major problems in astrophysics, such as the tides on compact object \citep{Piro2011}, Kilonova progenitors \citep{Metzger2019}, SN Ia progenitors \citep{Rebassa2019}.

A particular type of multimessenger source is the compact binaries containing an NS. NSs in the EM detectors can be found with radio and X-ray emission. Not only that, the orbital motion can emit low-frequency ($0.1\;\rm mHz-0.1\;\rm Hz$) GW signals for LISA, and the rapid rotation may emit high-frequency ($10-100\;\rm Hz$) GW signals for LVK observations. These distinctive properties enable us to proceed with dual-line detection of two-band GW signals simultaneously. The high-frequency GW signals of spinning NS arise from the ellipticities of NSs \citep{Abbott2017e,Tauris2018,Chenw2021}, accretion-built mountains \citep{Lasky2015,Haskell2017,Suvorov2021}, or other asymmetries that may produce mass quadrupole moments \citep{Cutler2002,Melatos2005,Lander2013,Andersson1999,Andersson2001,Andersson2014}. Unfortunately, such GW signals have not been captured in the LIGO O3 data \citep{Abbott2021b,Steltner2021,Stelner2023}. Maybe the upcoming next-generation GW detectors can answer this question: Do continuous GW signals exist for rapid rotation NSs?   

\section{Summary and conclusions}
\label{sec:7}
In this review, we have addressed the main formation scenarios of compact binaries as GW sources and discussed the physical properties of these sources in the GW observations. Substantial progress has been made both in single and binary stellar evolution, such as the final evolution of massive stars (e.g., \citealt{Heger2003,Janka2016}), binary mass transfer instability (e.g. \citealt{Chenx2008,Geh2010,Pavlovskii2015}), CE ejection prescription (e.g. \citealt{Hirai2022,Stefano2023,Geh2022}), etc. Nevertheless, there remain many uncertainties in the theoretical models, such as natal kick (e.g., \citealt{Janka2017,Muller2019}), CE ejection efficiency (e.g., \citealt{Scherbak2023,Geh2023b}), mass loss manners (e.g., \citealt{Luw2023,Picco2023}), etc. Future large-scale instruments, both EM and GW detectors, will enable us to better understand the fundamental problems in binary stellar evolution. Here, we point out some critical issues relating to the formation of compact binaries. 

\begin{itemize}
  \item The basic binary interaction processes. Binary mass transfer and CE evolution are binary evolution’s two most important physical processes. However, both processes produce some uncertainties, such as angular momentum loss manners, accretion efficiency, CE efficiency, etc. How to precisely constrain these processes is crucial to understanding the final remnants. The numerical simulations in high-dimension may hope to solve these fundamental problems of binary evolution. 
  \item The specific formation channels of accreting binaries. AM CVn and UCXBs are important target sources for future space-borne GW detectors. However, poorly understanding the formation channels makes the number predictions vary by one or two orders of magnitude. Therefore, the most pressing problem is to limit the formation channels based on the observation sample of AM CVn and UCXBs (e.g., \citealt{Ramsay2018,Armas2023}). What follows is the detailed binary population synthesis that considers all possible evolutionary routes.
  \item Individual formation channel for GW events. The stellar evolution processes in GW observations are still poorly constrained. The eccentricity is the most distinct characteristic in distinguishing the isolated binary evolution and dynamical formation models (or multiple interactions). However, due to the strong tidal circularize effect, only sources born with extremely large eccentricities can be observed in the LVK frequency band ($\gtrsim 10\;\rm Hz$; \citealt{Martinez2020}), and no such signals have been found so far. Another critical parameter in the GW observations is the BH spin. BBHs produced from different channels show various distributions of BBH spins (e.g., \citealt{Qiny2018,Marchant2023}). We expect further GW observations to improve our knowledge about the evolutionary scenarios. 
  \item Binary population synthesis on GW sources. Most works on the binary compact objects are based on the rapid population synthesis with considering the main binary interaction processes in a semi-analyzed way (see \citealt{Han2020} for a review). This method, without self-consistent binary evolution simulations, would result in large uncertainties. An improved method is to calculate large grids of models with detailed binary evolution code, which has been done in several works (e.g., \citealt{Fragos2023}). Binary population synthesis codes incorporating full stellar structure and binary evolution models are more complicated and time-consuming. Given the rapid development of numerical techniques and present-day computational power, we expect that the results from binary population synthesis can be more comprehensive.  
\end{itemize}

GW astronomy is a rapidly developing field, full of opportunities and challenges. We are fortunate to witness the era of GW astronomy, and we believe the next breakthrough in astronomy or basic physics will be made in the near future.

\section*{Acknowledgements}
{The authors thank the anonymous referees for helpful commments and suggestions.} We thank Dandan Wei for the meaningful discussions and comments on this review. We also thank Hailiang Chen, Hongwei Ge, Dengkai Jiang, Shuai Zha, Dongdong Liu and Chengyuan Wu for careful readings of the manuscript. {This work is supported by the Natural Science Foundation of China (grant Nos. 12125303, 12288102, 12090040/3, 11733008, 12103086, 11703081, 11422324, 12073070), the National Key R$\&$D Program of China (grant Nos. 2021YFA1600403, 2021YFA1600400), the Yunnan Revitalization Talent Support Program–Science $\&$ Technology Champion Project (No. 202305AB350003), the Key Research Program of Frontier Sciences of CAS (No. ZDBS-LY-7005), Yunnan Fundamental Research Projects (grant Nos. 202101AU070276, 202401AT070139), and the International Centre of Supernovae, Yunnan Key Laboratory (No. 202302AN360001). We also acknowledge the science research grant from the China Manned Space Project with Nos. CMS-CSST-2021-A10 and CMS-CSST-2021-A08.}




\bibliographystyle{elsarticle} 
\bibliography{main}






\end{document}